\documentclass{article}

\usepackage{
main}

\usepackage[utf8]{inputenc} 
\usepackage[T1]{fontenc}    
\usepackage{hyperref}       
\usepackage{url}            
\usepackage{booktabs}       
\usepackage{amsfonts}       
\usepackage{nicefrac}       
\usepackage{microtype}      
\usepackage{lipsum}
\usepackage{fancyhdr}       
\usepackage{graphicx}       
\usepackage{algorithm}
\usepackage{placeins}
\usepackage{algpseudocode}
\graphicspath{{media/}}     

\title{Design and simulation of memristor-based neural networks
}

\author{
  Pablo Alex Lázaro \\
  Universidad Complutense de Madrid\\
  \texttt{palex@ucm.es} \\
   \And
  Ignacio Jiménez Gallo \\
  Universidad Complutense de Madrid\\
  \texttt{ignaji01@ucm.es} \\
\And
Juan Roldán Aranda \\
Universidad de Granada\\
  \texttt{jroldan@ugr.es} \\
  \And
     Alberto del Barrio García \\
  Universidad Complutense de Madrid\\
  \texttt{abarriog@ucm.es} \\
\And
   Guillermo Botella Juan\\
  Universidad Complutense de Madrid\\
  \texttt{gbotella@ucm.es} \\
   \And    
     Francisco Jiménez Molinos\\
  Universidad de Granada\\
  \texttt{jmolinos@ugr.es} \\
}

\begin{document}
\maketitle

\begin{abstract}
In recent times, neural networks have been gaining increasing importance in fields such as pattern recognition and computer vision. However, their usage entails significant energy and hardware costs, limiting the domains in which this technology can be employed.

In this context, the feasibility of utilizing analog circuits based on memristors as efficient alternatives in neural network inference is being considered. Memristors stand out for their configurability and low power consumption.

To study the feasibility of using these circuits, a physical model has been adapted to accurately simulate the behavior of commercial memristors from KNOWM. Using this model, multiple neural networks have been designed and simulated, yielding highly satisfactory results.
\end{abstract}


\section{Introduction}
The field of machine learning and its applications in different areas, such as computer vision, natural language processing or robotics, is generating increasing interest in recent times. This has led to the development of advanced techniques and tools for the implementation of artificial neural networks. However, these digital neural networks require large amounts of energy and expensive compute resources, which limits their use in scenarios where energy efficiency is key.

Memristors have recently emerged as a promising alternative in the field of analog computation. Their memristive properties make them theoretically ideal building blocks for analog neural networks. These dedicated analog circuits, where the data and computation are combined forgoing the von Neumann architecture, are orders of magnitude more energy efficient and cheaper to manufacture.

\subsection{Motivation}
The training process of neural networks is an expensive process, both in terms of energy usage and hardware requirements, but it only needs to be done once. Inference, on the other hand, is performed every time the models are used. Many models in production are being used continuously and by many machines at once, so finding ways to reduce the energy and hardware expenses needed for inference is vital.

In some contexts, such as low power smart devices, the energy efficiency is a necessity. New breakthroughs in this field could bring new capabilities to these types of devices. 

There are already various techniques to achieve energy improvements, such as reducing precision and software optimizations, but the hardware constrains of the von Neumann architecture remain.

For many applications in industry, such as video encoding or cryptocurrency mining, single purpose accelerators are designed to speed up calculations and increase energy savings. Similarly, modern mobile chips come with small AI accelerators to perform matrix multiplications faster and consuming less power.

A memristor based analog neural network circuit could be employed in a similar manner, as a co-processor that would dramatically reduce the costs of running neural networks. The idea of using analog computation to perform inference on neural networks has already been explored in many studies, such as \cite{moreno2021cluster}

Memristors, with their unique properties, offer several advantages for analog computation over digital systems. Their non-volatility allows them to retain their state even when power is lost, making them ideal for memory storage. Inherently operating in an analog fashion, memristors allow for continuous data representation, which is more efficient for certain computations like neural networks, compared to the binary data representation in digital systems. Furthermore, memristors can be miniaturized and densely packed, providing more computational power in a smaller space. They also consume less power compared to digital transistors, enhancing energy efficiency. Perhaps most intriguingly, memristors can mimic the behavior of biological synapses, making them ideal for neuromorphic computing, a form of analog computation that emulates the brain's architecture. This paradigm is being the  subject of recent studies such as \cite{botella2009robust}

The main advantage of the memristors when compared to other analog alternatives is that they are both reconfigurable and non-volatile.

In summary, research on the viability of memristors in the field of neural networks is an important and current topic that can have significant implications in the development of computing and electronics.

\subsection{Objective}
The main objective of this Bachelor's Thesis is to design, simulate and compare inference results of analog neural networks based on memristors with equivalent transistor-based digital neural networks. We will use commercially available memristors to test whether they are a real alternative in their current state.

\subsection{Work plan}
To achieve the objective, the following steps will be taken:
\begin{enumerate}
    \item Study the properties and behavior of the device, using commercially available memristors from KNOWM and the Memristor Discovery software.
    \item Characterize the device by collecting real data, programming automated scripts and visualizing the results.
    \item Adjust a physical model of the memristor to faithfully represent our device.
    \item Design circuits to perform inference on analog neural networks, simulate and compare results.
\end{enumerate}   

\section{State of the Art}
\label{cap:estadoDeLaCuestion}

Next, we will explain the research perspective of the project carried out. The objective is to demonstrate an understanding of the theory necessary for the design of neural networks based on memristors.

\subsection{Introduction to Neural Networks}

Neural networks are computational models inspired by the human brain, they are widely used in the field of machine learning and can be described as universal function approximators.

These networks consist of interconnected nodes, called neurons, which work together to solve complex problems \cite{haykin1999neural}. The ability of neural networks to adapt makes them powerful tools in areas such as pattern recognition or natural language processing.

The behavior of these networks is based on the communication and processing of information through weighted connections between neurons. Each neuron receives multiple inputs, which are processed through a non-linear activation function to generate an output. The generated outputs are then sent as inputs to other neurons, creating a transmission and transformation chain of information throughout the network. In supervised learning, the weights of the neural networks are adjusted by comparing the outputs obtained with the desired results and using techniques such as gradient descent. 

This field has received a significant boost due to advances in computer technology and the availability of large volumes of data to train the models. In recent years, new types of neural networks have been developed, such as recurrent neural networks, convolutional neural networks, diffusion models, transformer networks and deep architectures \cite{goodfellow2016deep, bishop2006pattern}.

The importance of neural networks in the field of artificial intelligence lies in their ability to learn and generalize from data, enabling them to tackle complex problems and discover patterns in the data. This represents an advancement in various fields, including solutions to problems that are difficult to obtain using traditional algorithmic approaches.

The development of increasingly larger and more complex neural networks is driving the need for specialized hardware technologies that enable efficient implementations of these models.

In this context, a very promising line of research has emerged in the development of hardware for neural networks using memristors.

\subsection{Memristor}

Memristors are passive electric circuit components, such as inductors, capacitors and resistors. Their existence was theorized in 1971 by Professor Leon Chua in the article \cite{chua1971memristor}, titled: "Memristor-The Missing Circuit Element"  and they relate the electric charge and the magnetic flux together. There wasn't a physical implementation of them until very recently, when in 2008, a research team from Hewlett-Packard experimentally demonstrated the existence of these devices \cite{strukov2008missing}. 

\begin{figure}[h!]
	\centering
	\includegraphics[width = 0.25\textwidth]{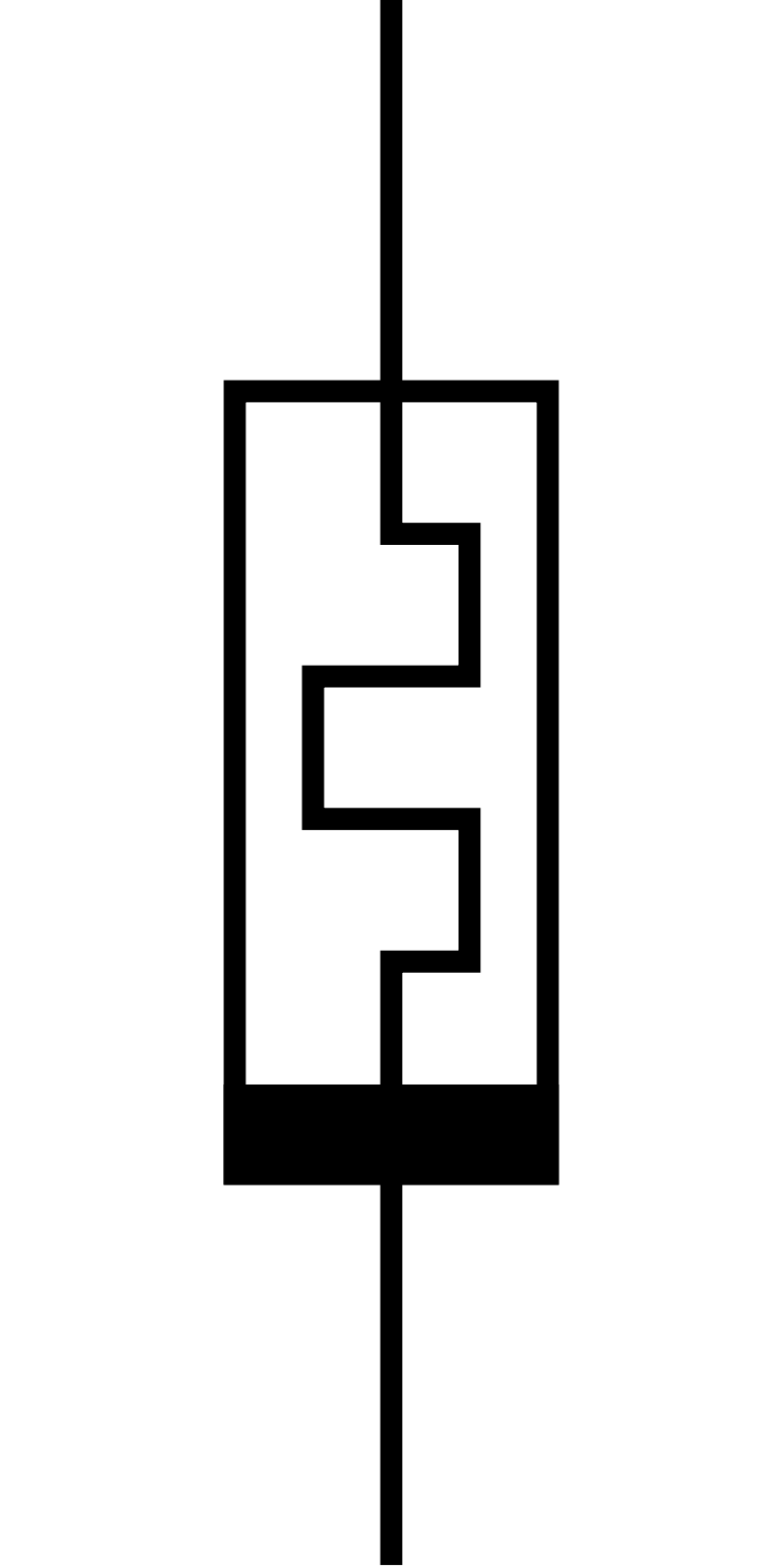}
	\caption{Memristor symbol}
\end{figure}

From a practical point of view, the memristor can be understood as a variable resistor whose resistance depends, in a nonlinear way, on the amount of current that has previously flowed through it. Additionally, it is a non-volatile device, meaning it retains its state even when disconnected from the current. These properties explain the name that Chua gave them: $memristor$, combining $memory$ and $resistor$.

This memory of its history is contained in its physical configuration, so it needs to be both:
\begin{enumerate}
    \item Physically reconfigurable when subjected to a certain voltage level. This reconfiguration must be reversible, going in both directions, from higher to lower resistivity when the voltage is positive and vice versa when it is negative.
    \item Stable, not changing its state once the current flow through it stops.
\end{enumerate}
These properties would exist in an ideal device; however, in practice, current implementations have the risk of reaching a physical configuration that becomes irreversible in some cases, such as a state of very high resistance. They also tend to slightly vary their state once disconnected from the current.
Information about this internal state of the memristor is captured in a state variable $\mu$. Its behavior is described by two equations:
\begin{enumerate}
\centering
\item$I(t) = G(\mu(t)) * V(t)$

\item$\frac{\mathrm{d}}{\mathrm{d}t}\mu(t) = F(\mu(t), V(t))$

\end{enumerate}
The first equation indicates that the current depends on the conductance (the inverse of resistance) that the memristor has in state $\mu$ at time $t$ and the voltage $V(t)$ applied at that moment. The second equation represents the change in the internal state of the device based on the current state $\mu$ and the voltage $V(t)$ it is being subjected to. These two equations give the device its non-linear behavior.

The operation of memristors is based on the phenomenon known as the "memristive effect." This effect occurs due to the properties of certain materials, such as metallic oxides, to change their resistance based on the direction and amount of electric charge passing through them. When a voltage is applied to a memristor, it undergoes a physical reconfiguration of its structure, changing its resistance, depending on the direction of the electric current.


\begin{figure}[h!]
	\centering
	\includegraphics[width = 0.4\textwidth]{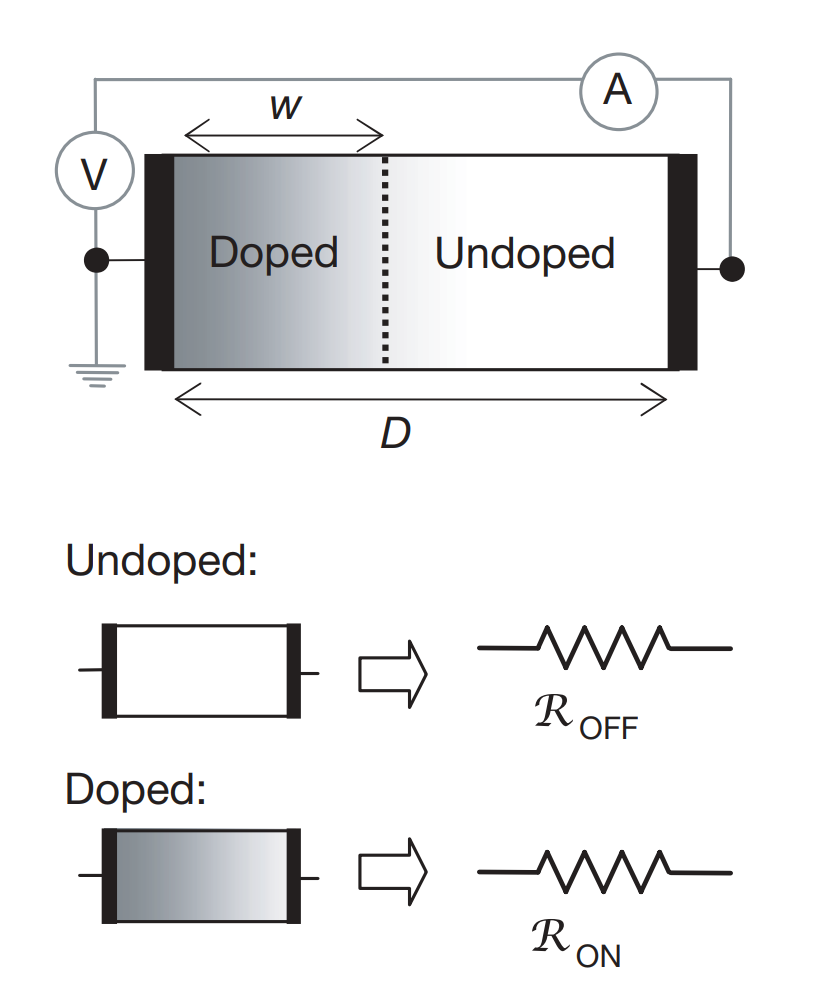}
	\caption{Simplified internal configuration of a memristor from \cite{strukov2008missing}}
\end{figure}

The ability to "remember" its previous resistance even after the applied voltage is removed is a key characteristic of memristors. This allows them to maintain their resistance despite the absence of an external power source, making them ideal candidates for the creation of artificial synapses in neural networks. \cite{yang2013science, pershin2011memory}.

Significant progress has been made in recent times in the fabrication of these devices, thus opening new possibilities in the field of neuromorphic computing and the design of hardware-efficient neural networks. \cite{indiveri2015memory}.

These devices have attracted great interest and study in artificial neural networks. This is because they offer potential advantages in terms of energy efficiency, storage density, and parallel processing, enabling the implementation of high-performance neural networks. \cite{pershin2011memory}.

Memristors can emulate the behavior of artificial synapses. Synapses are essential elements in neural networks as they facilitate communication between neurons and information processing. Traditionally, these neural synapses have been implemented using transistors and capacitors, but this emerging technology offers a promising alternative \cite{yang2022research, hu2018memristor}.

Most of the simulations performed on the papers that explore this idea, like in \cite{HASAN201731}, use resistors or very rough models to approximate the behaviour of memristors. We will use state of the art physical models in our simulations, accounting for the inherent variability of the device.

\section{Memristor Discovery}  \label{Memristor Discovery}

\subsection{Calibration}

In this first process, the focus was on getting acquainted with memristor technology. To achieve this, the Memristor Discovery pack from KNOWM was used, which includes a chip with eight pairs of memristors, a board with assigned resistors, and a manual for the software application. To utilize this tool, the logic analyzer of the Analog Discovery 2 from Digilent was employed.

\begin{figure}[htbp]
    \centering
    \begin{minipage}[b]{0.4\linewidth}
        \includegraphics[width=\linewidth]{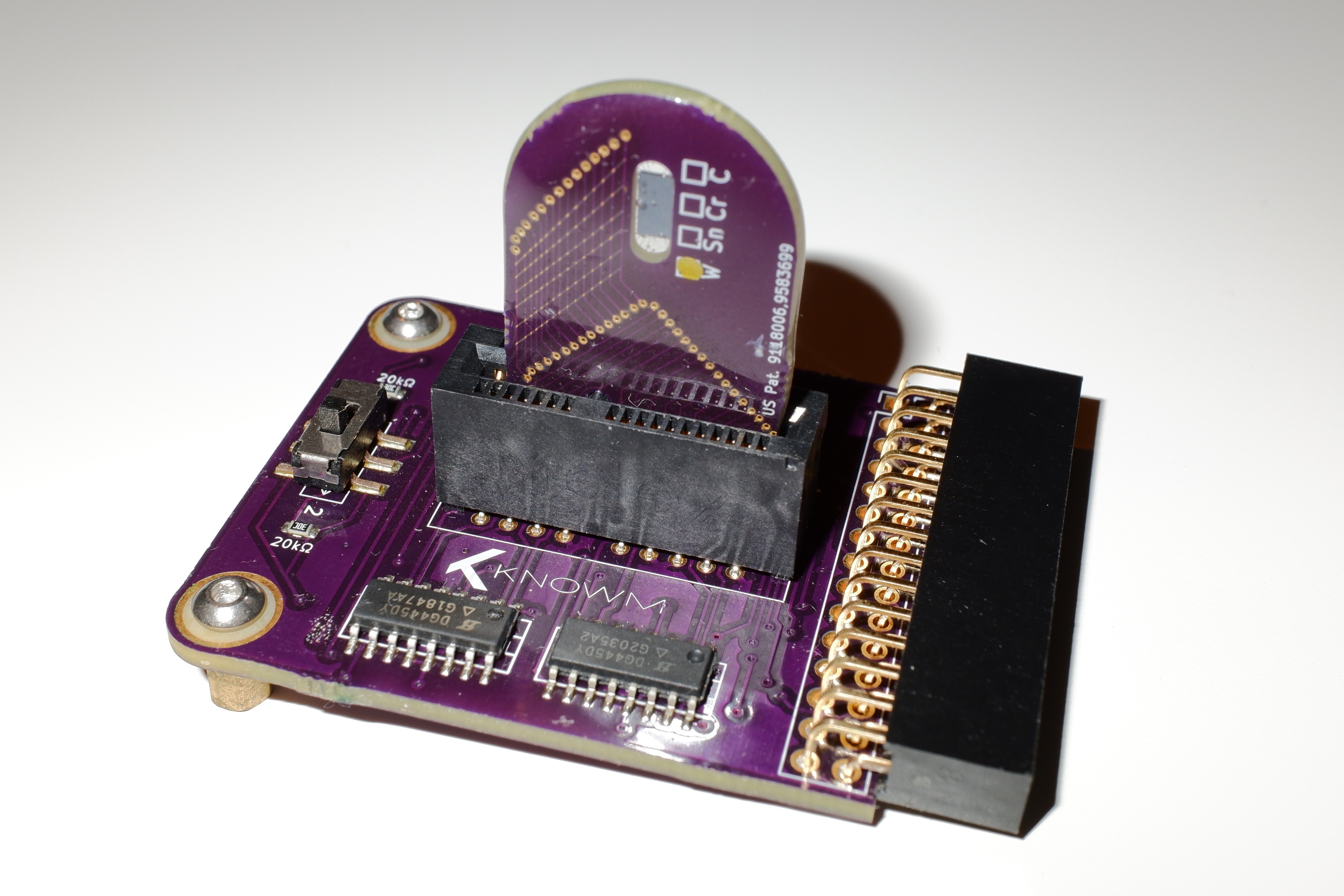}
        \centering
        {Memristor Discovery}
    \end{minipage}
    \hspace{0.1\linewidth}
    \begin{minipage}[b]{0.4\linewidth}
        \includegraphics[width=\linewidth]{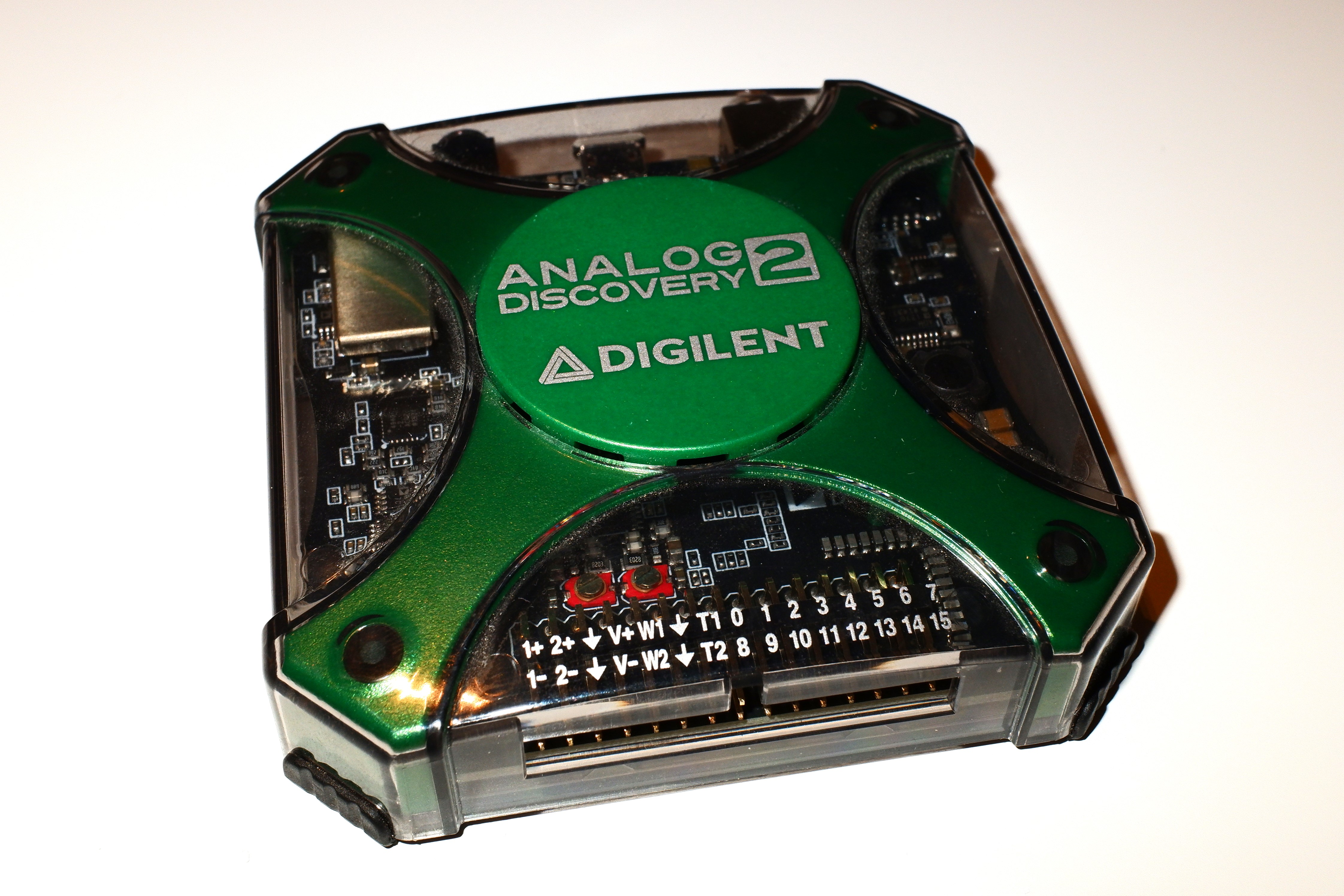}
        \centering
        {Analog Discovery 2}
    \end{minipage}
    \caption{Memristor Discovery and Analog Discovery 2}
\end{figure}

The setup of the Analog Discovery 2 using the Waveforms software was initiated. Firstly, the logic analyzer was connected to a laptop via USB. Once connected, an automatic test of wave generation functions was performed. To do this, the thirty-pin connector was used, connecting waveform generator 1 to oscilloscope 1 and waveform generator 2 to oscilloscope 2. Next, the software was configured to generate a square wave and a sinusoidal wave, as shown in figure 3.2.

\begin{figure}[h!]
	\centering
	\includegraphics[width = 0.6\textwidth]{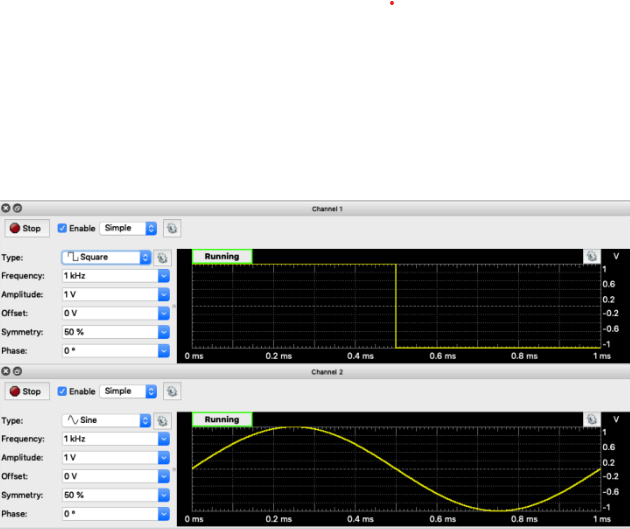}
	\caption{Waveform generators configured for self-test.}
\end{figure}

Next, the logic analyzer was calibrated to ensure accurate measurements. A multimeter was used to measure the voltage generated at each step of the calibration process to correct the generated signals.

With the calibrated logic analyzer and the memristors functioning properly, we can now begin to conduct our own experiments using KNOWM's memristors through the Memristor Discovery application. In order to familiarize ourselves with this hardware, we will perform various tests, ranging from obtaining hysteresis curves to designing our own simple neural network.

\subsection{DC Experiment }

First, we started by exploring the DC option of the KNOWM Memristor Discovery software. This functionality allows us to characterize the electrical properties of a memristor easily and efficiently.  So, to initiate our DC experiment, we apply a DC voltage to a memristor and then measure the resulting current. The current-voltage (I-V) and conductance-voltage (G-V) graphs are then plotted to show the memristor's characteristic pinched hysteresis behavior.

The following figures show the I-V and G-V graphs for a memristor driven with an input sinusoidal waveform. The I-V graph shows that the memristor has a non-linear resistance, also appreciated in the G-V graph that shows how conductance (G) does not vary proportionally to the applied voltage (V). The hysteresis loop shows that the memristor has a memory effect, meaning that its resistance depends on its previous state.
\begin{figure}[h!]
	\centering
	\includegraphics[width =\textwidth]{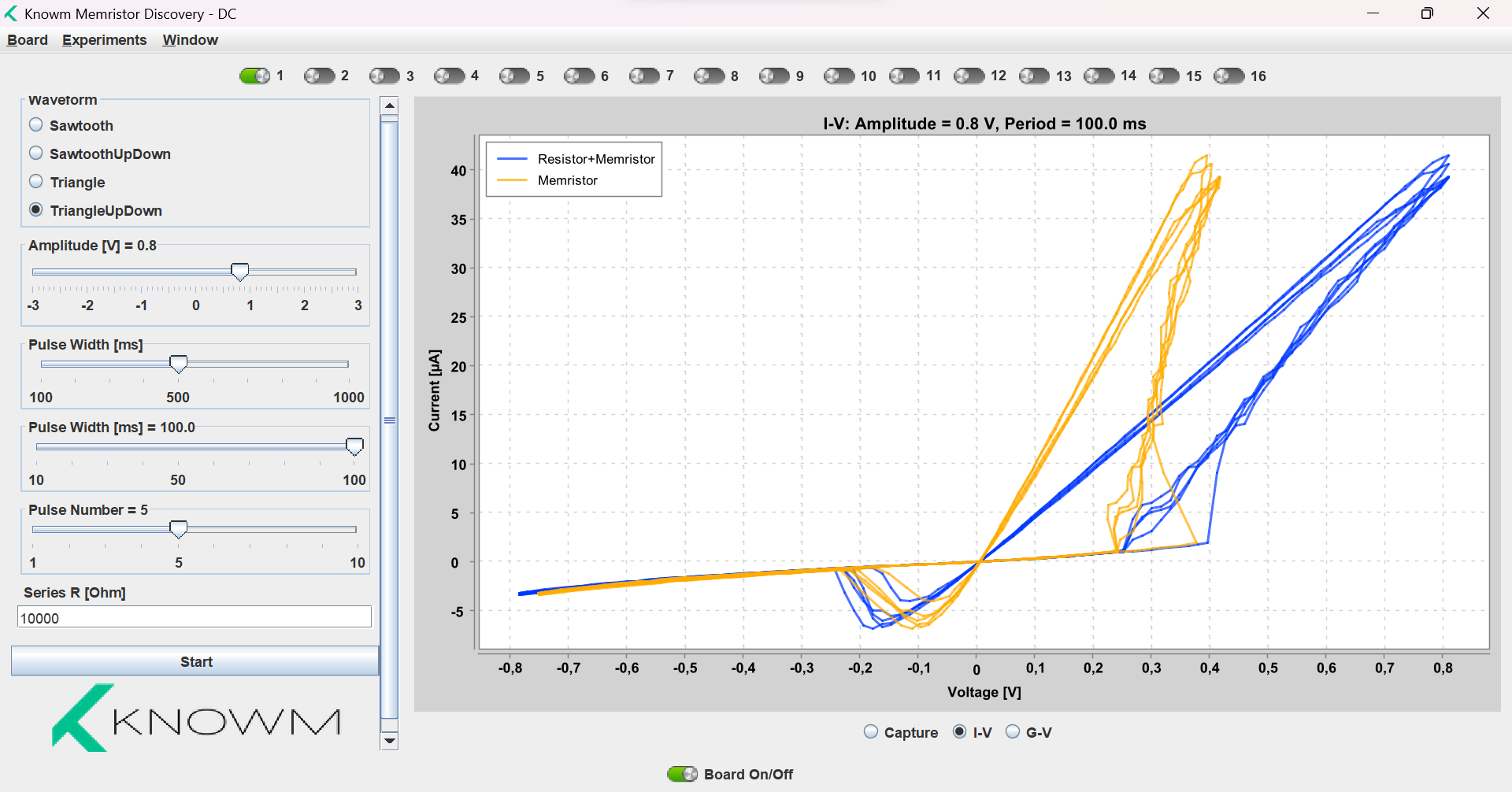}
	\caption{DC Response I-V.}
 \label{DC Response I-V}
\end{figure}
\begin{figure}[h!]
	\centering
	\includegraphics[width =\textwidth]{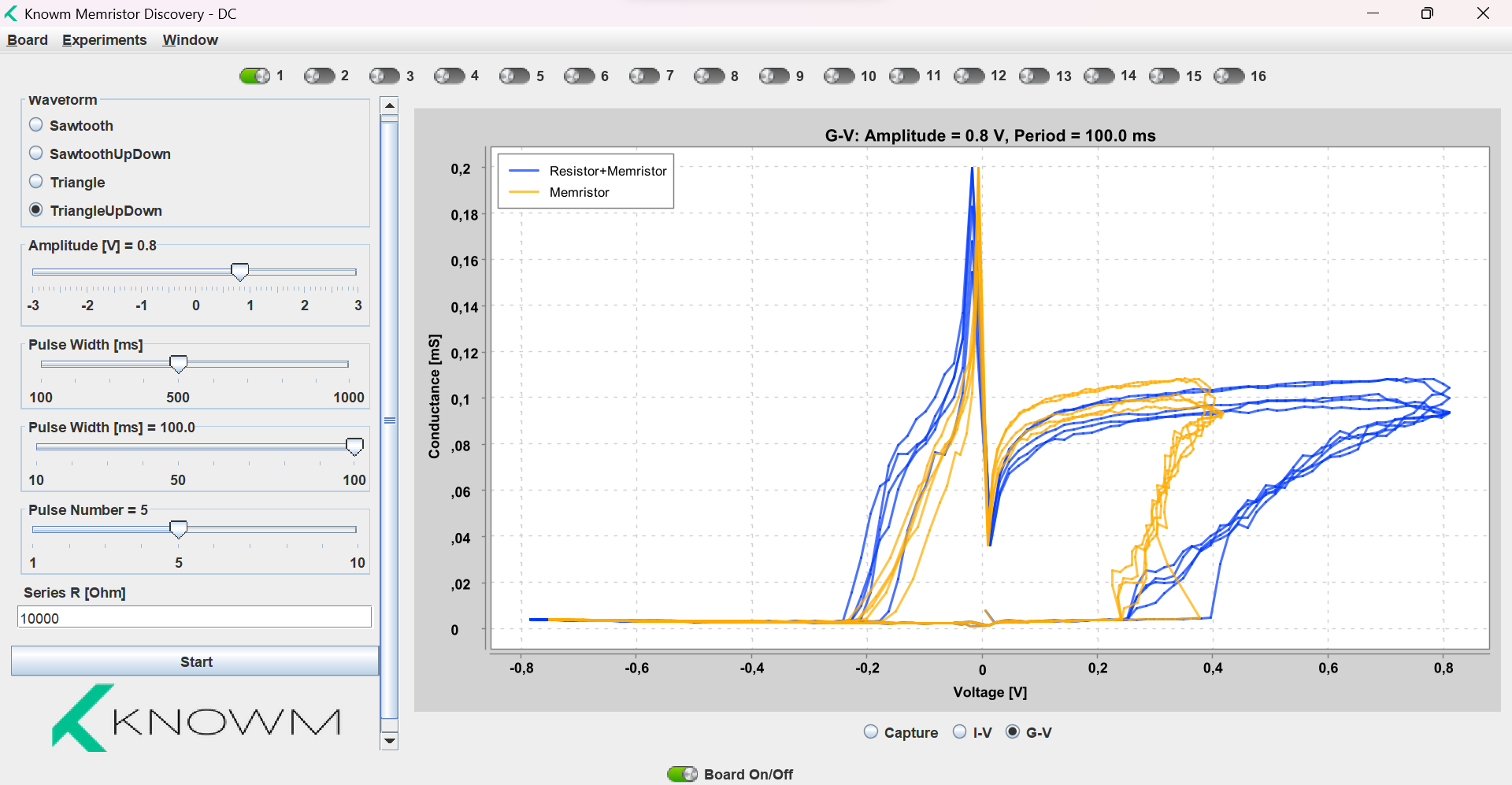}
	\caption{DC Response G-V.}
 \label{DC Response G-V}
\end{figure}

This DC experiment is a powerful tool for characterizing memristors. It can be used to measure a variety of important parameters, including the resistance, capacitance, switching threshold, and non-linearity. This information can be used to design circuits that take advantage of the memristor's unique properties.

It is worth noting that the yellow-colored graphs represent the data obtained if the resistor were not applied, while the blue-colored graphs represent the behavior of the memristor in conjunction with the resistor.

\subsection{AC Experiment }

Next, we proceed to investigate the possibilities of the Hysteresis option within the same software. To do this, we will create a new experiment. In it, a memristor is driven by an AC voltage source and the resulting current is measured. The current-voltage (I-V) graphs are then plotted to show the memristor's characteristic pinched hysteresis behavior.

\begin{figure}[h!]
  \centering
  \includegraphics[width=0.75\linewidth]{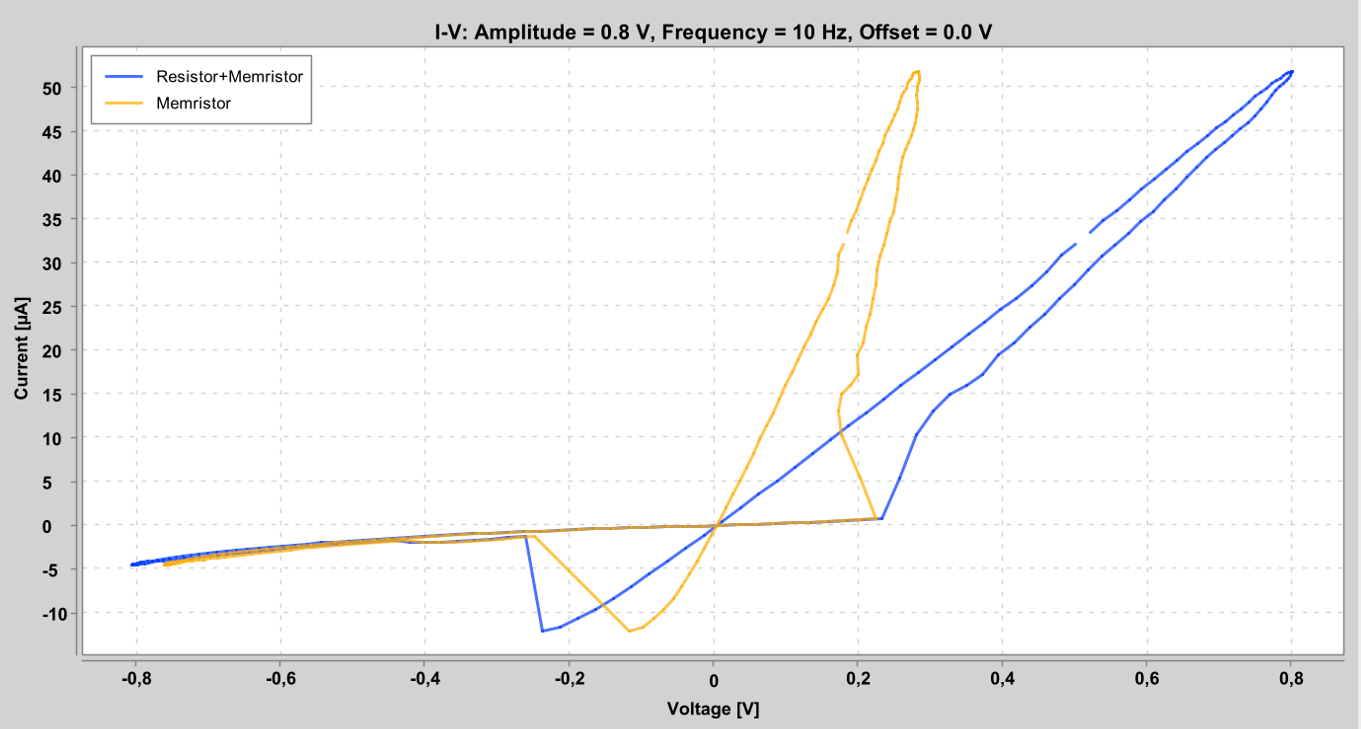}

  \caption{AC Experiment 10Hz}
\label{AC Experiment 10Hz}
\end{figure}

\begin{figure}[h!]
  \centering
  \includegraphics[width=0.75\linewidth]{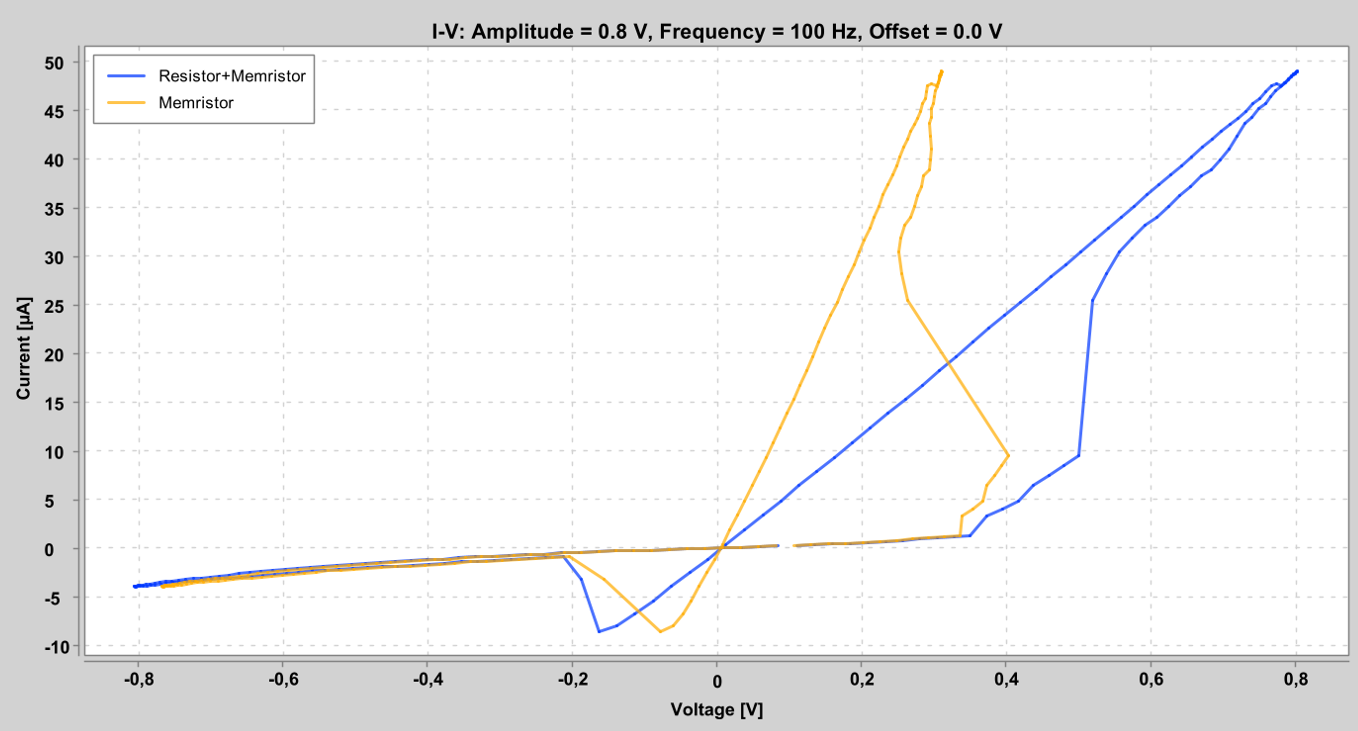}

  \caption{AC Experiment 100Hz}
\label{AC Experiment 100Hz}
\end{figure}

  \begin{figure}[h!]
  \centering
  \includegraphics[width=0.75\linewidth]{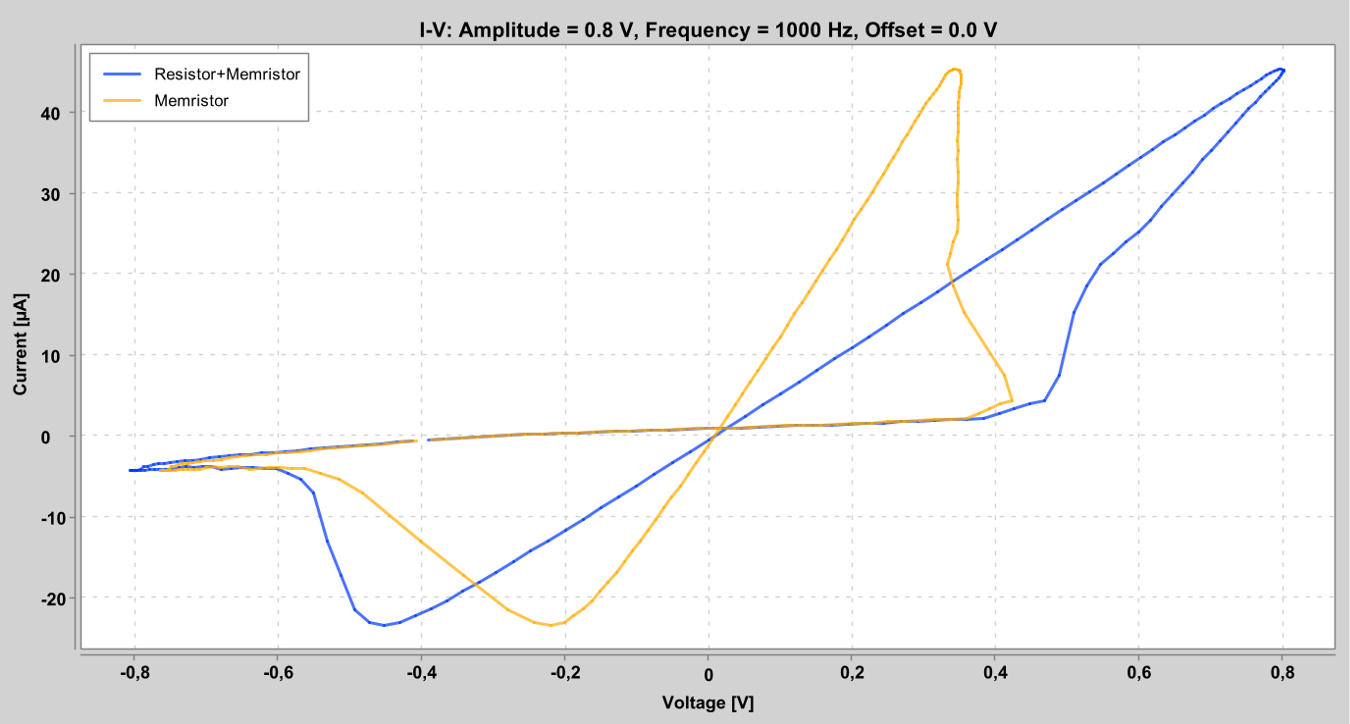}

  \caption{AC Experiment 1000Hz}
\label{AC Experiment 1000Hz}
\end{figure}

The frequency of the AC voltage source affects the memristor's hysteresis behavior. To test it, we applied three types of ramps, that is, three different frequencies. These frequencies were 10Hz, 100Hz, and 1000Hz. At low frequencies the memristor has a larger hysteresis loop. This is because the memristor has time to switch between resistance states at low frequencies. While  at high frequencies, the memristor has a smaller hysteresis loop, because the memristor spends less on either state and more time switching between the resistance states. This can be verified by comparing the different plots in Figures \ref{AC Experiment 10Hz}, \ref{AC Experiment 100Hz}, \ref{AC Experiment 1000Hz}, where it can be observed that the hysteresis curve for a frequency of 1000Hz is flatter than the curve for 10Hz.

\subsection{Resistance Programing}

Continuing with the option of Hysteresis, a new experiment was conducted. In this experiment, a memristor is driven by a DC voltage source and the resulting current is measured. The current is then used to calculate the memristor's resistance. The memristor's resistance can be programmed by changing the DC voltage source.

This Resistance Programming experiment works by applying a DC voltage to the memristor. The DC voltage causes the memristor to switch between two resistance states. The resistance state that the memristor switches to depends on the polarity of the DC voltage.

The Resistance Programming experiment can be used to program the resistance of a memristor in a variety of ways. One way to program the resistance is to apply a DC voltage to the memristor for a specific amount of time. The longer the DC voltage is applied, the larger the change in the memristor's resistance.
\begin{figure}[h!]
	\centering
	\includegraphics[width =0.75\textwidth]{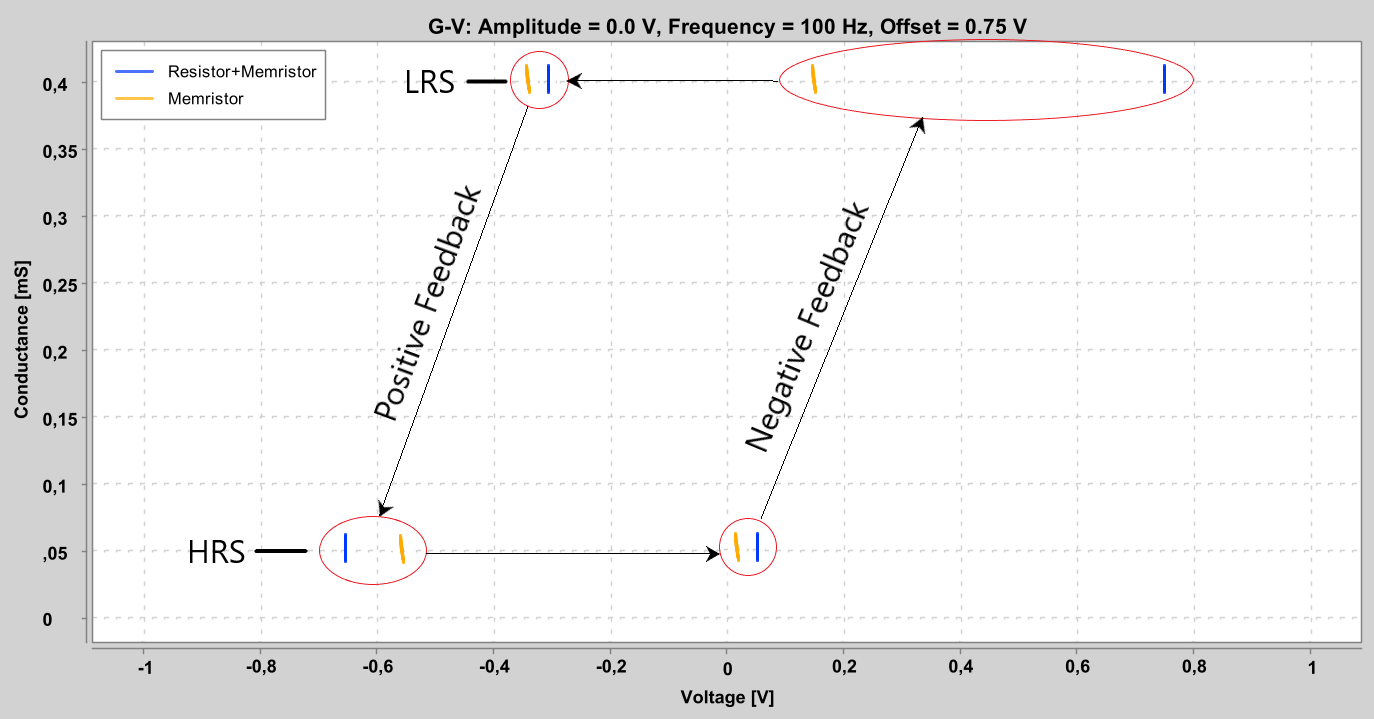}
	\caption{Resistance Programming}
\end{figure}
The HRS and LRS refer to the high resistance state and the low resistance state, respectively. The HRS is the state in which the memristor has a high resistance, and the LRS is the state in which the memristor has a low resistance. The memristor can be programmed to switch between the HRS and LRS by applying a voltage to the memristor.

\subsection{Pulse Response }

We continue exploring the Pulse option. For this experiment, a memristor is driven with a series of pulses of varying amplitude and width. The response of the memristor to these pulses is then measured and analyzed.
\begin{figure}[h!]
	\centering
	\includegraphics[width =\textwidth]{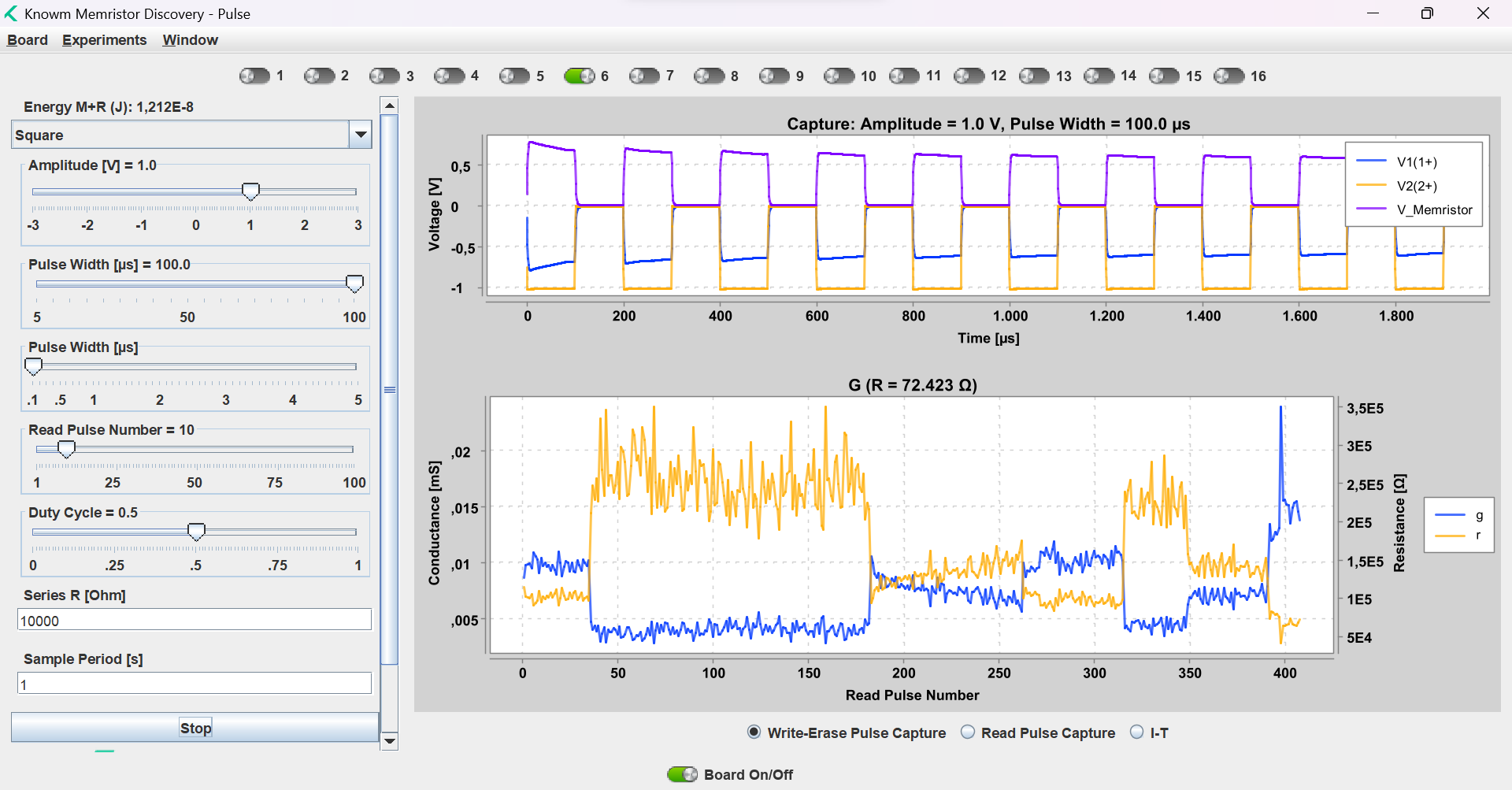}
	\caption{Pulse Response}
 \label{Pulse Response}
\end{figure}

The graph shows at image \ref{Pulse Response} the relationship between the conductance and resistance of a memristor as a function of the number of read pulses applied to the memristor. The graph shows that the conductance of the memristor increases with each pulse, while the resistance decreases. This is due to the fact that the memristor is a non-volatile memory device, meaning that it can store information even after the power is removed.

\subsection{Synapse Experiment}

 This experiment was run in Synapse12 option of the KNOWM software. It consists of a series of elemental kT-RAM instructions that are used to drive the synapses, followed by a series of FLV read instructions that are used to measure the synaptic state and synaptic pair conductances. This experiment is repeated multiple times to observe the continuous response of the synapses.

\begin{figure}[h!]
	\centering
	\includegraphics[width =\textwidth]{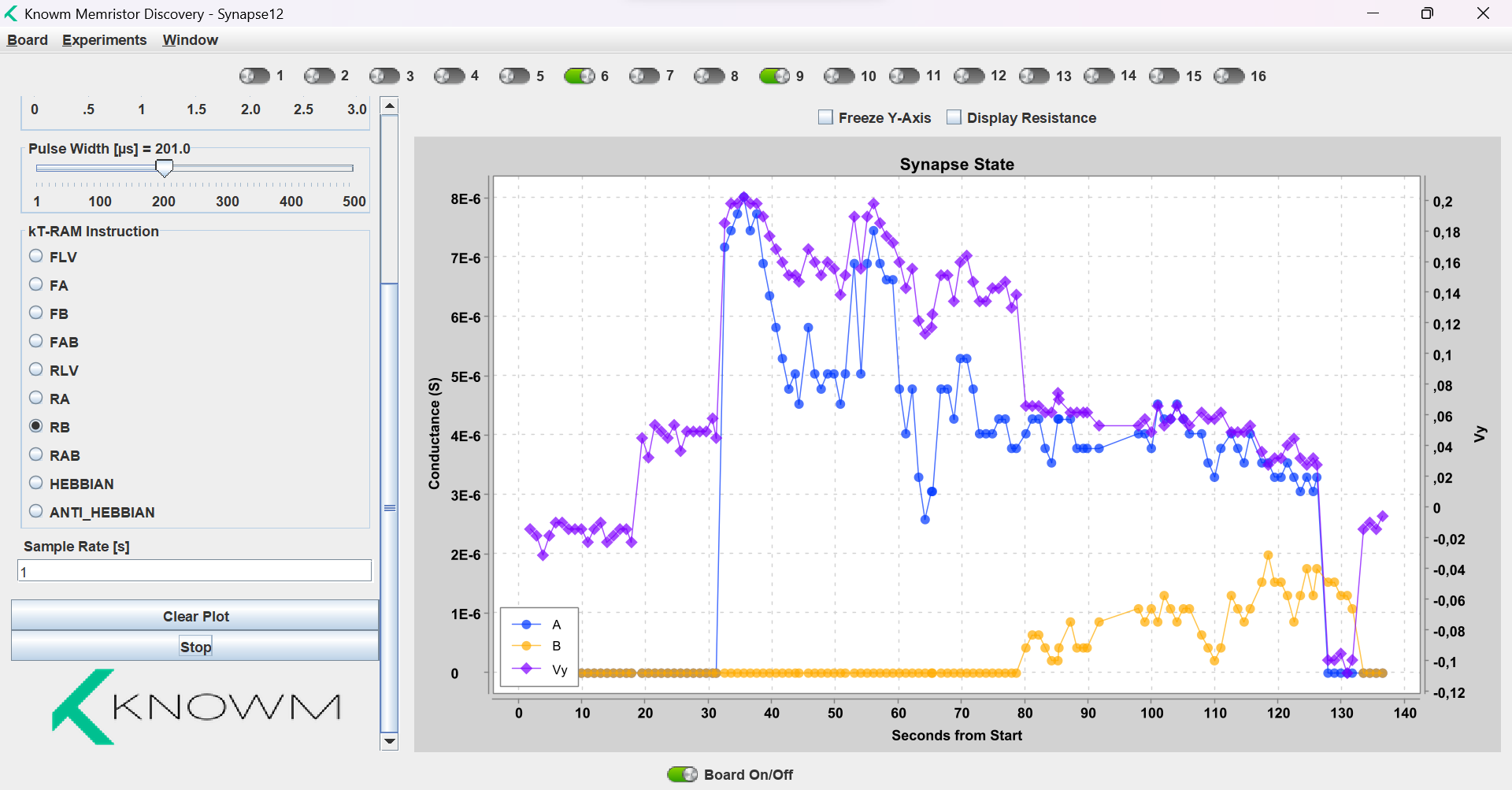}
	\caption{Synapse Experiment}
 \label{Synapse Experiment}
\end{figure}

The significance of this experiment is that it provides a detailed characterization of the functionality of kT-RAM differential-pair memristor synapses. The figure \ref{Synapse Experiment}  demonstrates that the synapses can be driven with a variety of elemental instructions, and that the synaptic state and synaptic pair conductances can be accurately measured with FLV read instructions. This experiment also demonstrates that the synapses exhibit a continuous response, which is a key requirement for their use in neuromorphic computing applications.

\subsection{Classifier Experiment}

Finally, we started experimenting with the Classify12 option, that is a software program that allows users to test the performance of a memristor-based classifier. The program uses a 1X16 linear array chip to create an 8-synapse neuron. The user can select the forward and reverse pulse voltage amplitudes and pulse widths for each synapse.

So, we create a Classifier Experiment that is significant because it allows us to test the performance of a memristor-based classifier without having to design our own hardware.

\begin{table}[h!]
\centering
\begin{tabular}{||c c c||} 
 \hline
 Dataset & True Patterns & False Patterns\\ [0.5ex] 
 \hline\hline
 Ortho2Pattern & [0,1,2,3] & [4,5,6,7] \\ 
 \hline
 AntiOrtho2Pattern & [4,5,6,7] & [0,1,2,3]\\
 \hline
 Ortho4Pattern & [4,5],[6,7] &[0,1],[2,3] \\
 \hline
 AntiOrtho4Pattern & [0,1],[2,3] & [4,5],[6,7] \\
 \hline
 Ortho8Pattern & [4],[5],[6],[7] & [0],[1],[2],[3]\\ 
 \hline
 AntiOrtho8Pattern & [0],[1],[2],[3] & [4],[5],[6],[7]\\ 
 \hline
\end{tabular}
\caption{Classifier Experiment Datasets}
\label{Classifier Experiment Datasets}
\end{table}

To use the Classifier Experiment, we first need to select the forward and reverse pulse voltage amplitudes and pulse widths for each synapse. Afterwards, we select one of the training data set shown in table \ref{Classifier Experiment Datasets} . The training data set should consist of a set of input and output pairs. The input pairs are the patterns that the classifier will be trained to recognize. The output pairs are the desired outputs for each input pattern.

When the training data set has been selected, we start the training process. The training process will take several iterations. After each iteration, the classifier will be updated to improve its performance. The training process will stop when the classifier has reached the desired epoch.

Once the training process is completed, without resetting the weights obtained in the memristor pairs, we run it with a new data set. The new data set should  be used for training. The classifier's performance on the new data set shows how well the classifier will fit a new target despite having been trained before.

\begin{figure}[h!]
	\centering
	\includegraphics[width =\textwidth]{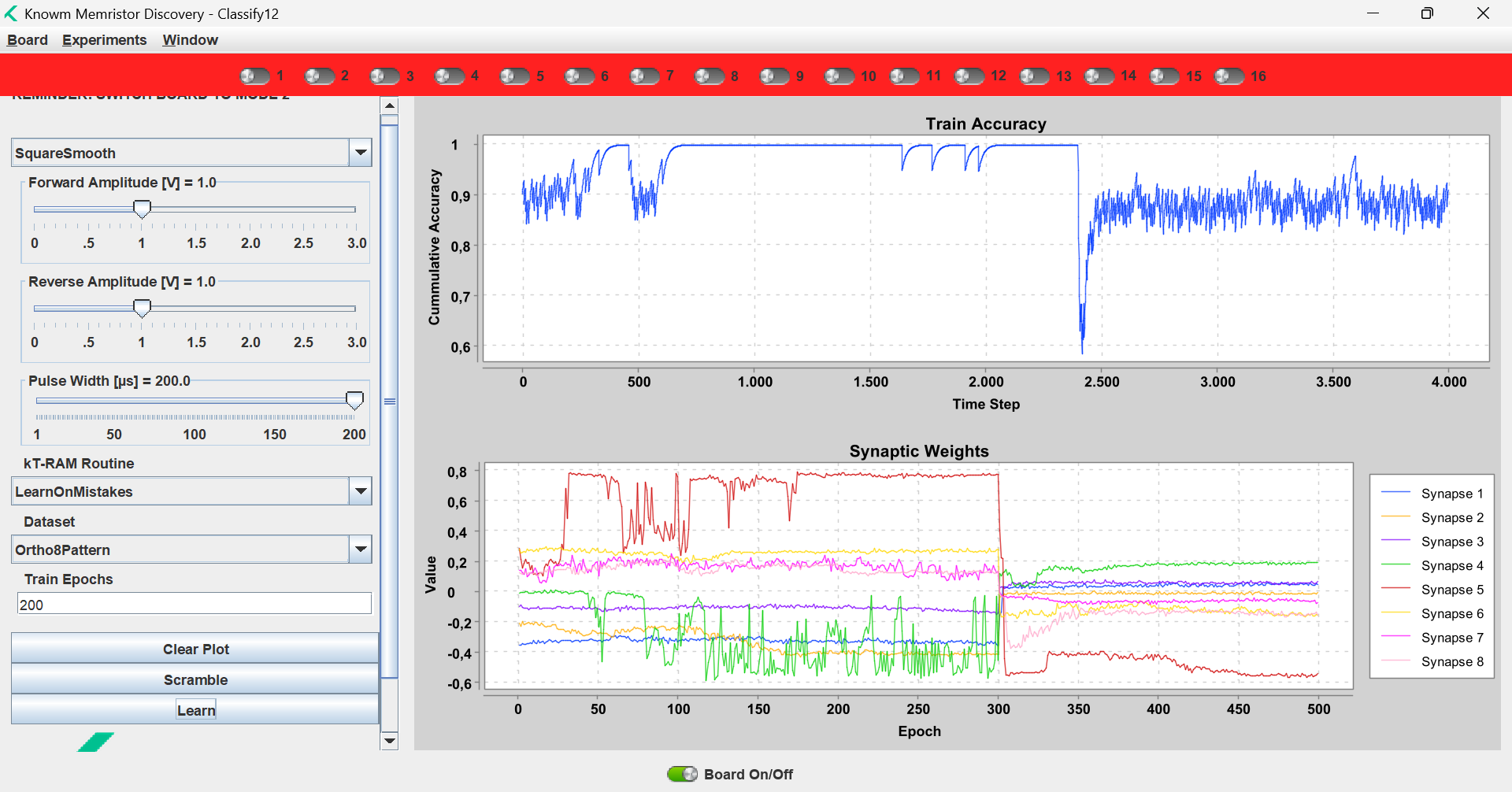}
	\caption{Resistance Programming}
 \label{fig:clasificacion_binaria}
\end{figure}
In Figure \ref{fig:clasificacion_binaria}, you can see how the classification is produced using the Ortho8Pattern dataset, where synapses 4, 5, 6, and 7 had to classify as true (positive), while the rest of the synapses had to classify as false (negative). Shortly after starting the execution, you can observe how it reaches 100\% accuracy. Without resetting the resistance weights, the dataset is changed to AntiOrtho8Pattern, which should invert the values to be classified, and it can be observed how all synapses modify their weights, obtaining an accuracy close to 90\%.

To delve deeper into the functioning of memristors as synapses, we designed a simple classification experiment by creating both a new dataset shown in table \ref{Custom Dataset Memristor} and a new learning method. We wrote our code with the extensions in Java in the source code of the program. 

\begin{table}[h!]
\centering
\begin{tabular}{||c c c||} 
 \hline
 Dataset & True Patterns & False Patterns\\ [0.5ex] 
 \hline\hline
 myDataSet & [0, 2, 4], [2, 4, 6] & [1, 3, 5], [3, 5, 7] \\ 
 \hline
\end{tabular}
\caption{Custom Dataset Memristor}
\label{Custom Dataset Memristor}
\end{table}

The learning method called LearnComboReverse consists of learning only when it must classify as false, or when it fails.

\begin{figure}[h!]
	\centering
	\includegraphics[width =\textwidth]{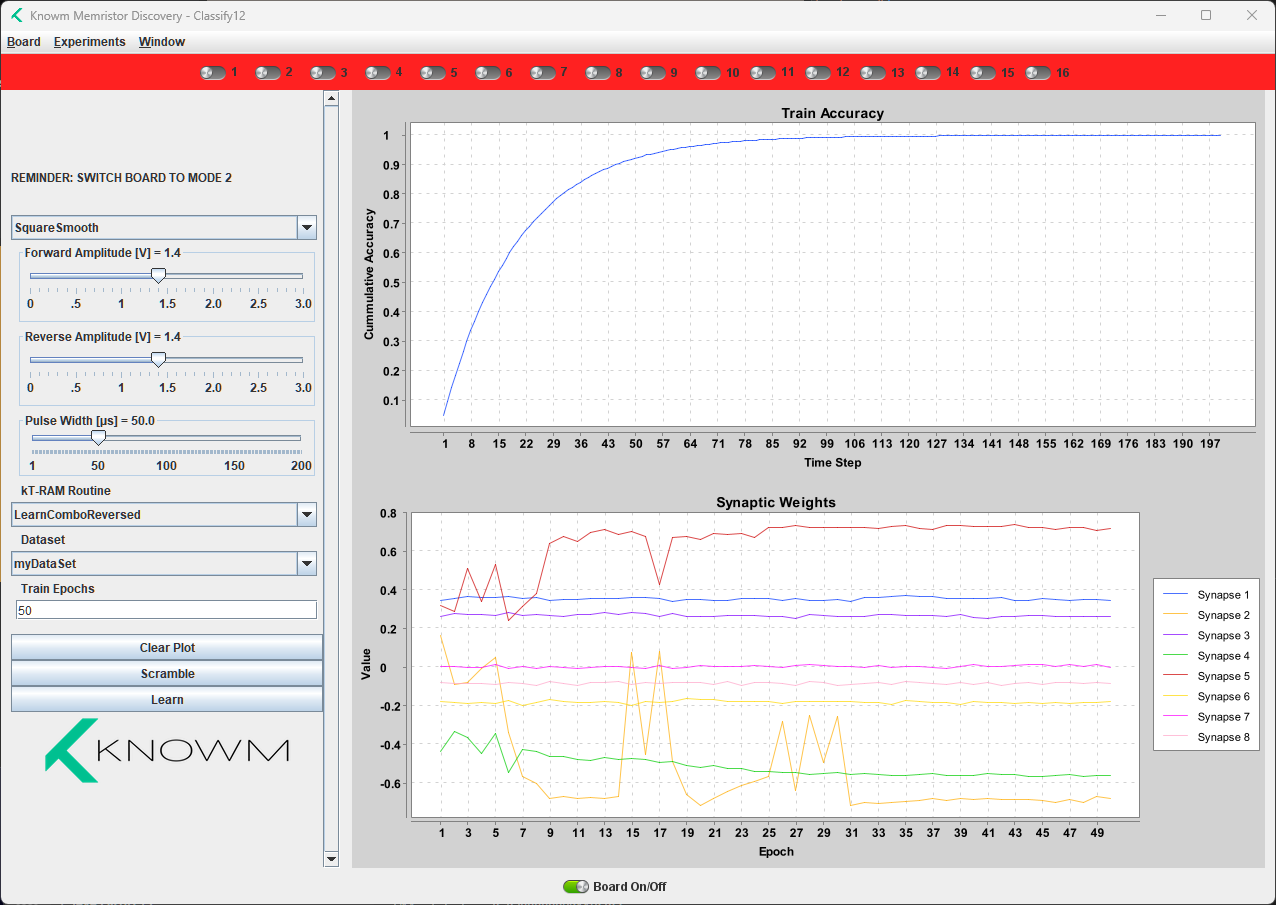}
	\caption{Custom Classification}
\end{figure}

\section{Device Characterization}
\label{cap:devicecharacterization}
Once the experimentation phase was done and an intuitive understanding of the device had been achieved, we set out to experimentally characterize it. This consists of obtaining the numerical data needed to determine some of the devices’ most important properties, such as the set and reset voltages and the HRS and LRS currents ratio for various voltage ramps. These results are obtained by means of automated extraction scripts that are executed on the experimental data we collected from the device.

\subsection{Data Collection} \label{Data Collection}
The DC Experiment environment from Knowm Memristor Discovery program provided us with the perfect means to visualize and obtain the data we needed. We first scanned through a few of the sixteen memristors on the chip until we selected one that showed good and consistent behavior. For that memristor, we applied $200$ pulses of ramp voltage stress (triangle shaped voltages) with an amplitude of $0.8 V$ through its $10K\Omega$ resistor; and we did so for three different pulse periods: $10ms$, $100ms$ and $1000ms$ , which result in ramps of $0.8V/s$, $8V/s$ and $80V/s$ respectively.

\begin{figure}[h!]
	\centering
	\includegraphics[width = 0.7\textwidth]{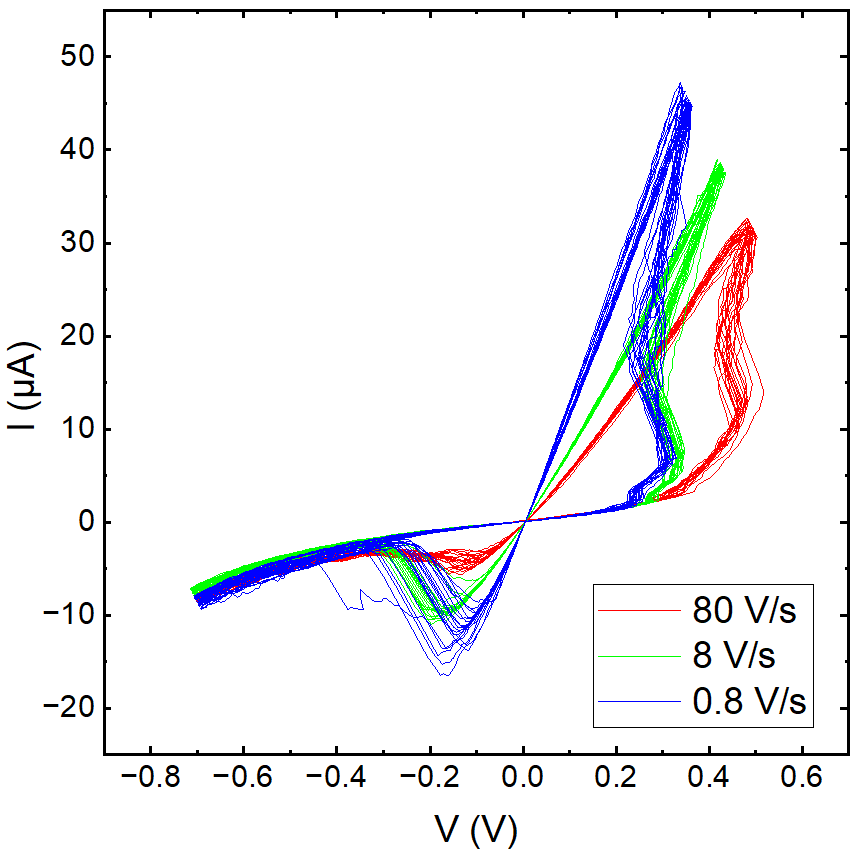}
	\caption{Measured current for the applied voltage ramps}
\end{figure}

The program only allowed us to apply ten pulses at a time, so we slightly modify its source code to see if we could obtain the two hundred cycles in a single execution, eliminating the downtime involved in saving the data and rerunning the execution every time. We thought this downtime could affect the results due to the non-ideal nature of our devices, resulting in noticeable jumps in the measurements due to the tendency of our memristors to lose some conductance when idle. After examining the source code of the Knowm Memristor Discovery and Waveforms programs, both written fully in Java, we came across a hardware buffer size limitation on the Discovery board itself, where the results were stored before being sent via USB to the computer. We managed to double the  number of cycles to 20 but couldn't go any further. The effects we feared are slightly noticeable in some of the results, but for the most part, it didn't present a real issue.

\subsection{Data and script adaptation} \label{Data and script adaptation}
The first properties we obtained were the set and reset voltages of the device. These are the points at which the device starts to meaningfully transition from the HRS to the LRS for the set voltage and vice-versa for the reset voltage. This information is interesting since it provides us with an interval $(V_{reset} - V_{set})$ of the voltages we can apply to the device expecting to not see any meaningful changes in its internal state.

To obtain these two points, we had to find the parameters that would better met our specific needs using a MATLAB script for the methods from \cite{MALDONADO2022111876}. We required the input data to be separated by the set and reset of each cycle and for the currents to be in absolute values. For that purpose, we developed a Python script.

\begin{figure}[h!]
	\centering
	\includegraphics[width = 0.9\textwidth]{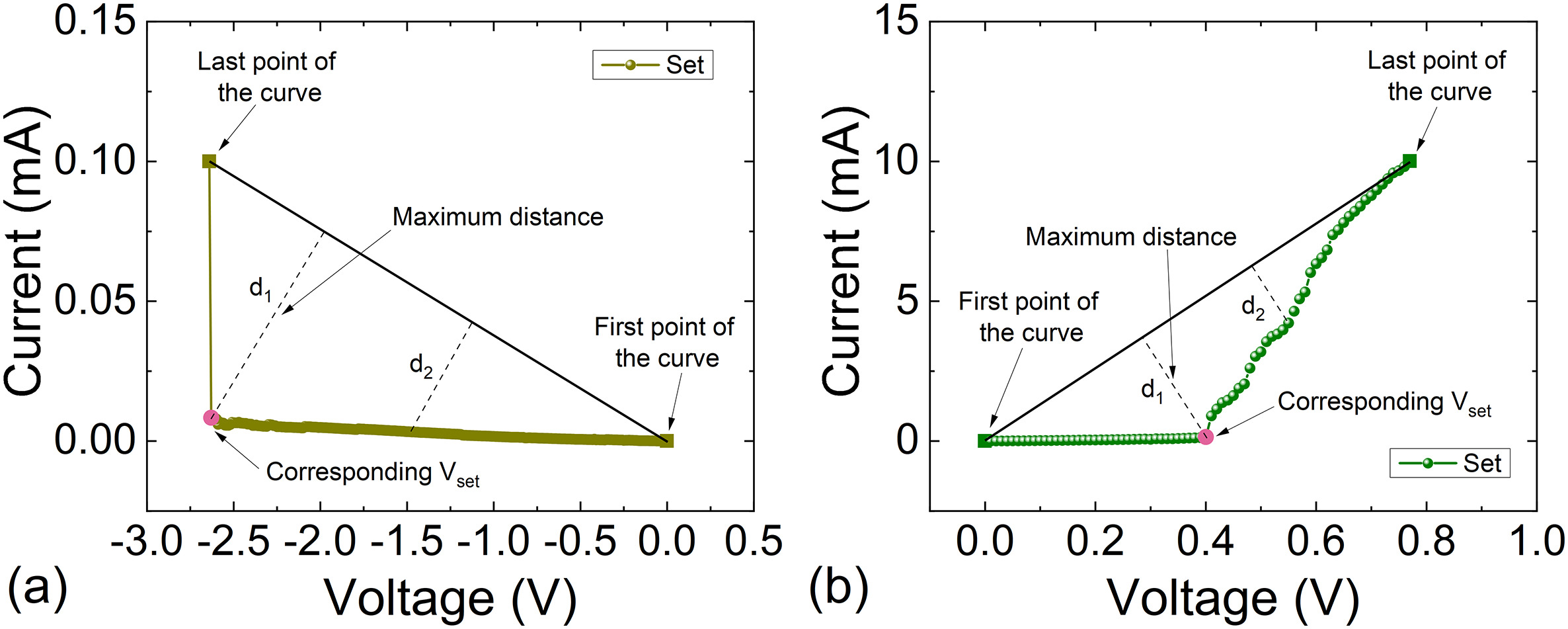}
\caption{Method for selecting the set point based on maximum separation}\
    \label{method3}
\end{figure}

Each method provided us with different results for each cycle. We selected the ones that better represent the experimental measurements, which were the third method from the paper for both the set and reset. For the set, the method selected the points where there was a maximum separation from the straight line that joined the first and end points of each set curve, as shown in \ref{method3}. For the reset, the method simply selected the points with the highest current.

We were also interested in showing the cycle to cycle variability at a given voltage points, and more precisely, at a reading voltage point, such as $0.05V$ or $0.1V$ in both HRS and LRS. For this purpose, we had to write another python script to estimate the current at these points using a linear interpolation on the real data since we didn't have the values for those exact voltages. 

\subsection{Results}
\label{sec:char_results}
The results presented here, are soon to be published in a paper named "Characterization and modeling of variability in commercial self-directed channel memristors" in the 14th Spanish Conference on Electron Devices (CDE 2023). \cite{cde2023characterization}.

\begin{figure}[h!]
    \centering
    \begin{minipage}{0.5\textwidth} 
          \centering
          a)
          \includegraphics[width=\linewidth]{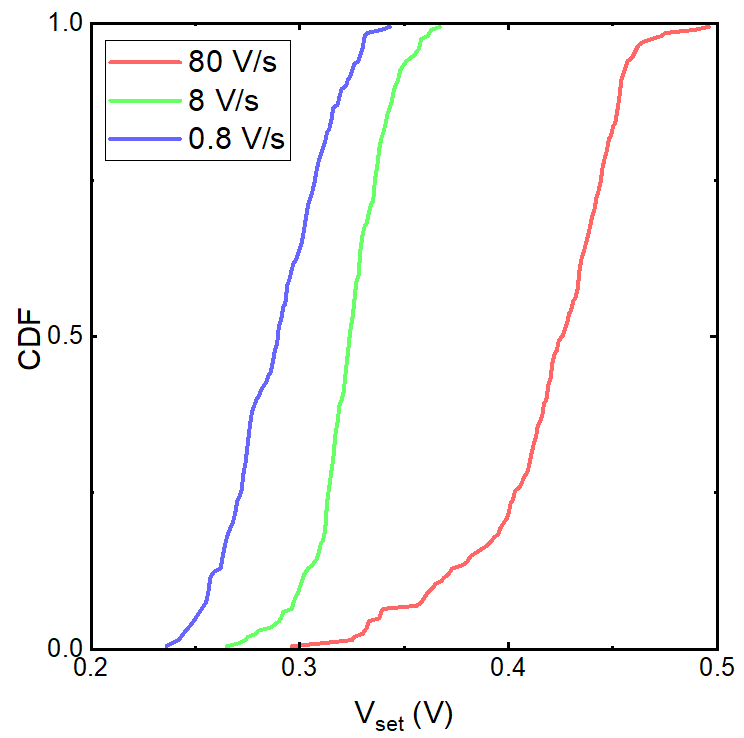}
    \end{minipage}%
    \begin{minipage}{0.5\textwidth} 
          \centering
          b)
          \includegraphics[width=\linewidth]{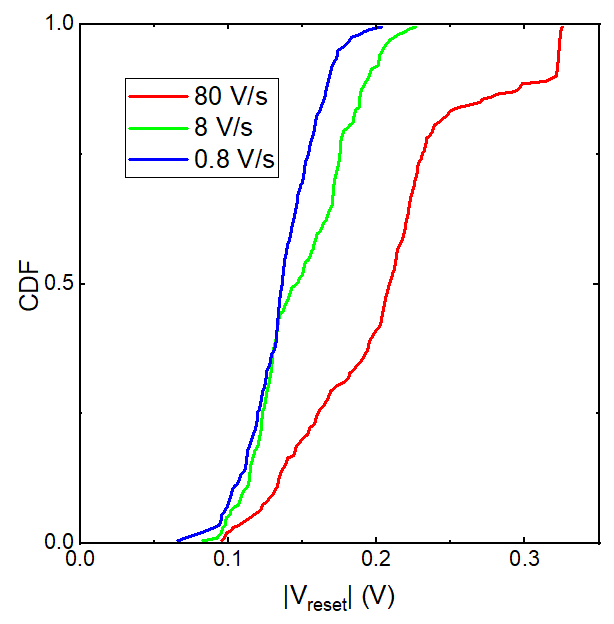}
    \end{minipage}
     \caption{Cumulative distribution functions of the calculated set (a) and reset (b) voltages}
     \label{CDFs}
\end{figure}

We chose to use a cumulative distribution function (CDF) to illustrate both the voltage set and reset of the device across all the $200$ cycles. The results, seen in figure \ref{CDFs},  are fairly consistent and in line with the manufacturer's advice to use $0.1V$ as probing voltage. The resulting interval of voltages in which the device can operate without meaningfully changing its resistance is roughly from $-0.1V$ to $0.25V$. This information will be useful when designing the neural networks.

\begin{figure}[h!]
	\centering
	\includegraphics[width=0.5\textwidth]{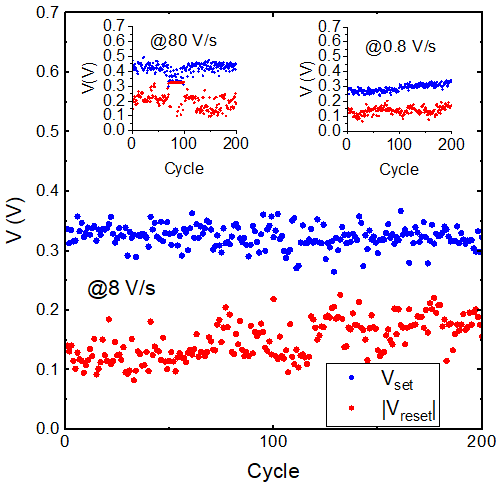}
	\caption{Set and reset voltages versus cycle number, for the three voltage ramp rates}
    \label{cycle_points}
\end{figure}

It is also interesting to see in figure \ref{cycle_points} how this voltage points vary from cycle to cycle. Although a slight general tendency can be intuited, the variability seems to be consistent across all voltage ramps.

\begin{figure}[h!]
      \centering
      \includegraphics[width=0.5\linewidth]{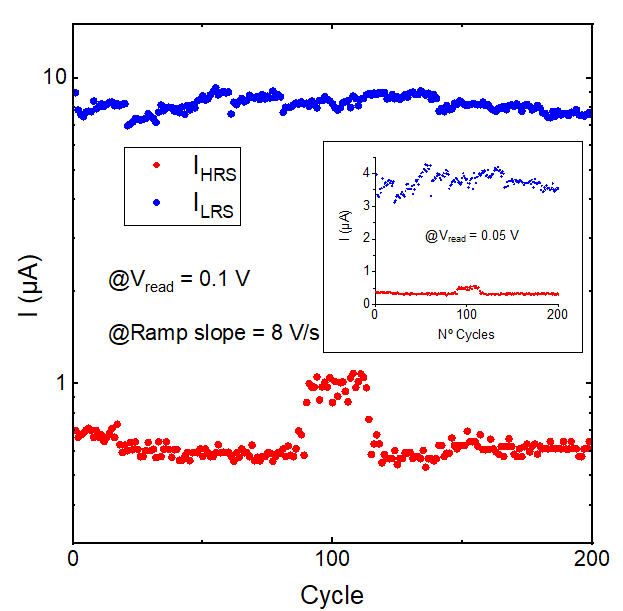}
      \caption{Cycle to cycle variability of the current measured at 0.1V in HRS and LRS}
      \label{c2c}
\end{figure}

\begin{figure}[h!]
      \centering
      \includegraphics[width=0.5\linewidth]{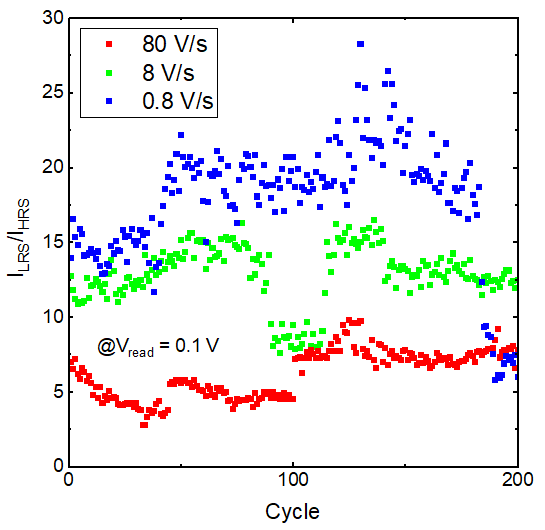}
      \caption{figure}{HRS and LRS measured currents ratio versus cycle number for 0.1V}
      \label{ratios}
\end{figure}

Lastly, the current estimated at $0.1V$ and $0.05V$ at HRS and LRS can be seen in figure \ref{c2c} for the $8V/s$ measurements. Except for some cycles around cycle $100$, they appear very consistent. However, when we plot the ratios of currents in LHR and HRS for $0.1V$, we can see much more variability (figure \ref{ratios}). There are a few reasons for this, firstly we have to take into account the significant measurement noise when dealing with these small currents and the fact that interpolation was performed to obtain these values. We also think that in this case some of the variability can be attributed to the interrupted measurement process that we were forced to do, since some of the jumps appear to coincide exactly with multiples of $20$.

\FloatBarrier

\section{Physical Model}
\label{chap:physical_model}

Due to the limitation of having only sixteen memristors for the creation of minimally complex neural networks such as MNIST, a physical model has been created in LTspice to emulate the behavior of memristors.

\begin{figure}[h!]
	\centering
	\includegraphics[width =\textwidth]{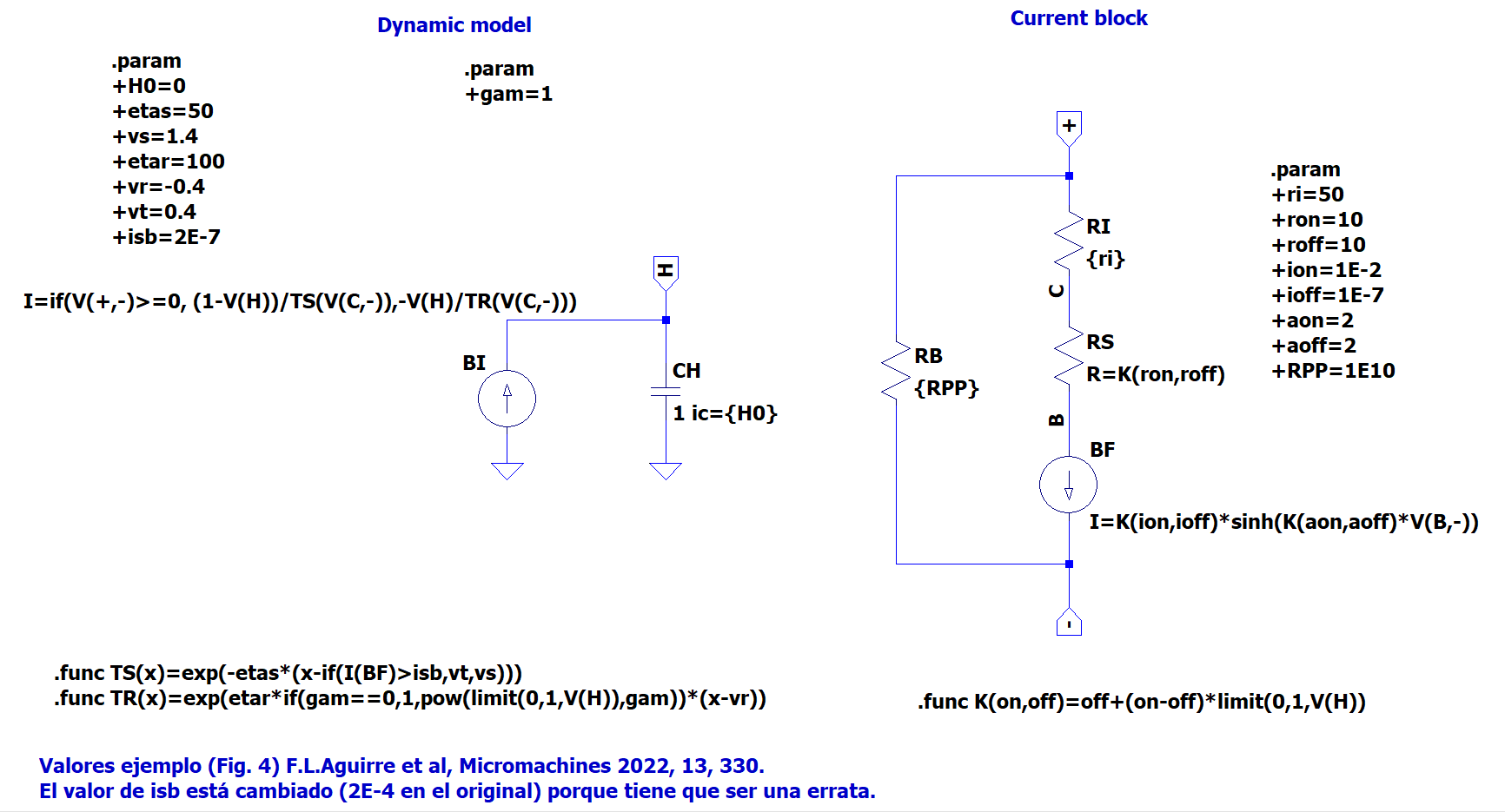}
	\caption{Memristor Physical Model}
 \label{Memristor Physical Model}
\end{figure}

\begin{figure}[h!]
	\centering
	\includegraphics[width =0.8\textwidth]{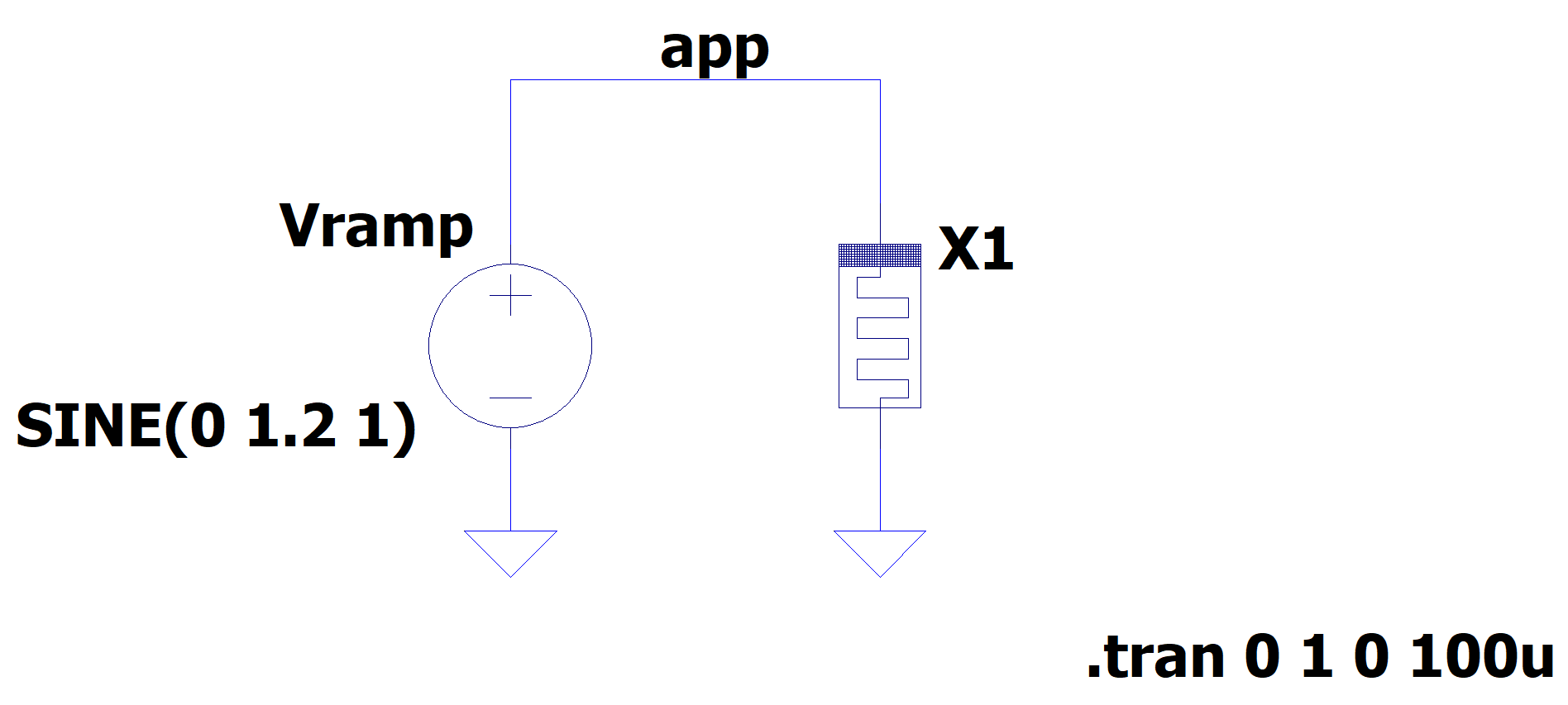}
	\caption{Physical Model}
 \label{Physical Model}
\end{figure}

\begin{figure}[h!]
	\centering
	\includegraphics[width =0.7\textwidth]{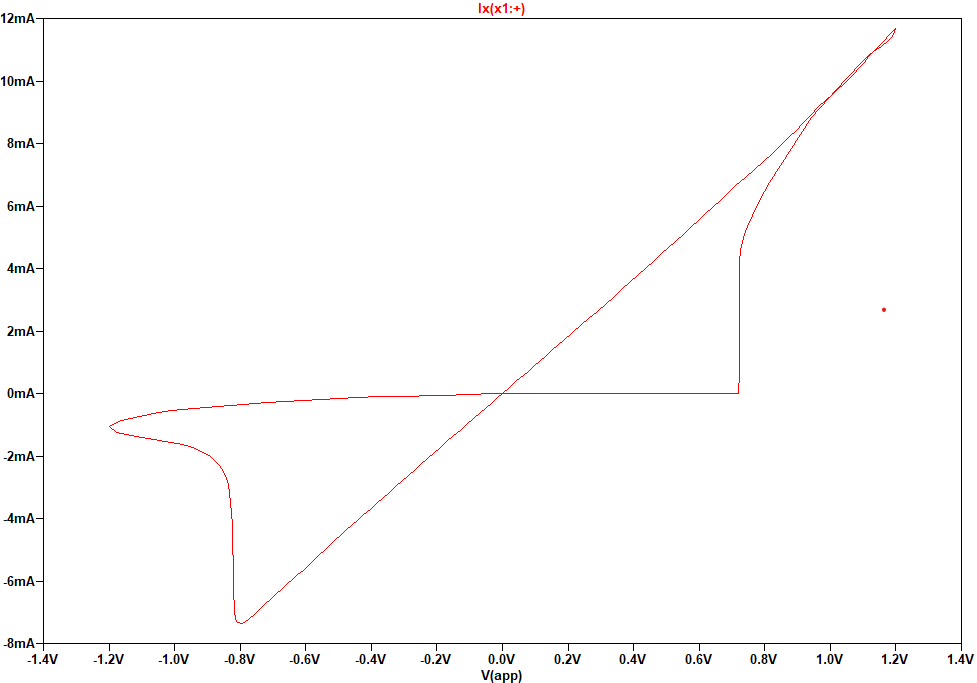}
	\caption{LTspice simulation}
 \label{LTspice simulation}
\end{figure}

To do this, we relied on the parameters described in \cite{Aguirre202214}. As seen in the model shown in Figure \ref{Memristor Physical Model} and \ref{Physical Model}, we applied a voltage of 1.2V to the memristor, obtaining the response shown in Figure \ref{LTspice simulation} using the LTspice simulator, graphing a curve very similar to the characteristic hysteresis curve of memristors on the I - V(app) axes.

However, the model is not yet complete as commercial memristors exhibit a certain level of variability despite having stochastic components. This fact causes them to behave differently during two different cycles and leads to different memristors functioning differently despite being under the same conditions.

Due to this situation, we have emulated this variability behavior within our model. To do so, we have added the parameters mentioned in the paper \cite{Roldan202333}. In order to appreciate the variability, unlike the previous model, we execute the model a given number of times, in the case of Figure \ref{LTspice simulation with variability}, one hundred times. It can be clearly observed that the result is very different from our empirical data. However, this will be addressed by using a genetic algorithm to adjust the initial parameters as to match our devices as accurately as possible.

To perform a simulation with variability, we use the Gaussian function 'gauss' in LTspice. This function generates a random number from a Gaussian distribution with a standard deviation determined by the parameter used. The function is applied to the new parameters that serve to emulate variability.

\begin{figure}[h!]
	\centering
	\includegraphics[width =\textwidth]{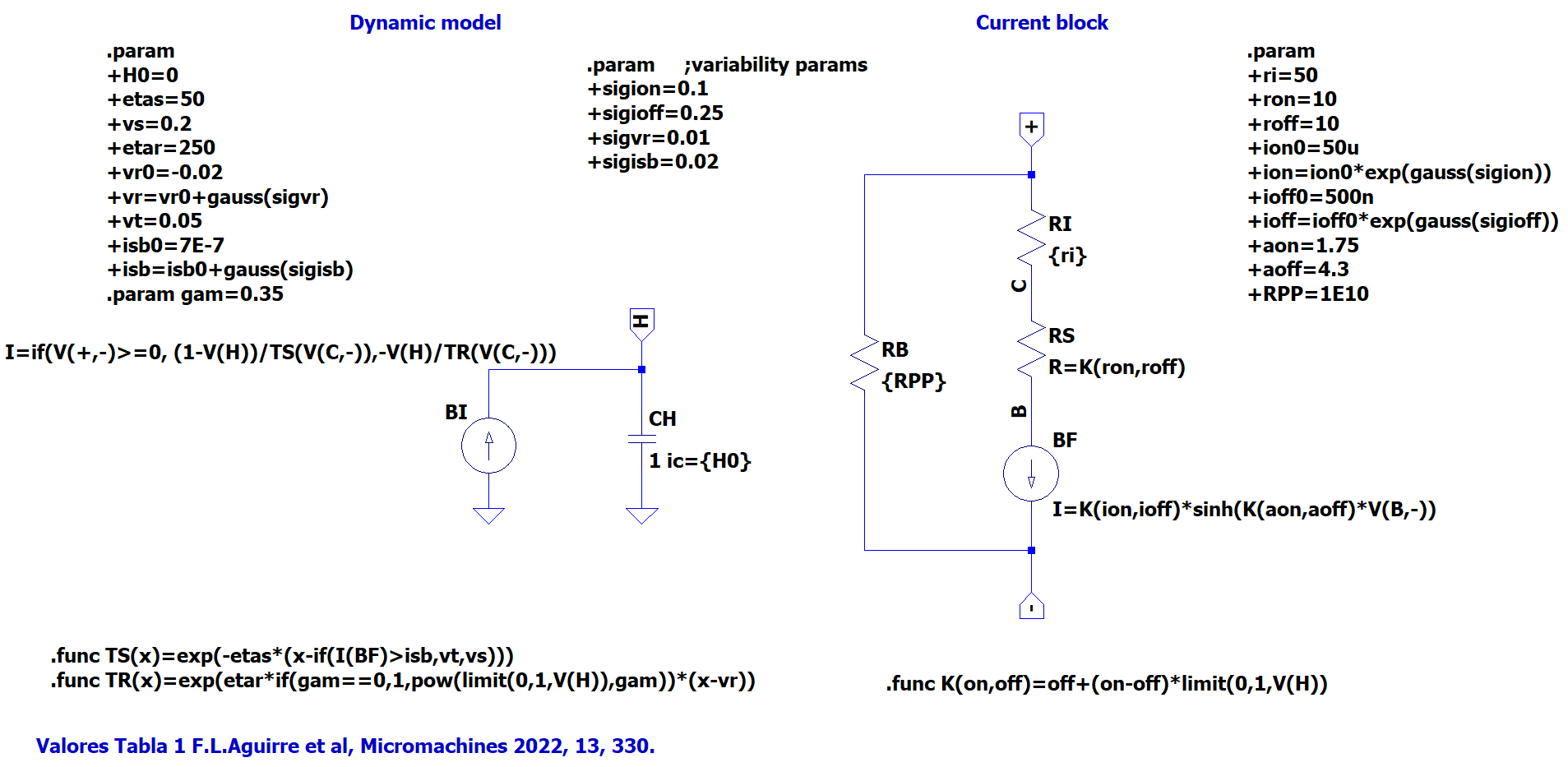}
	\caption{Memristor Physical Model with variability}
\end{figure}

\begin{figure}[h!]
	\centering
	\includegraphics[width =0.7\textwidth]{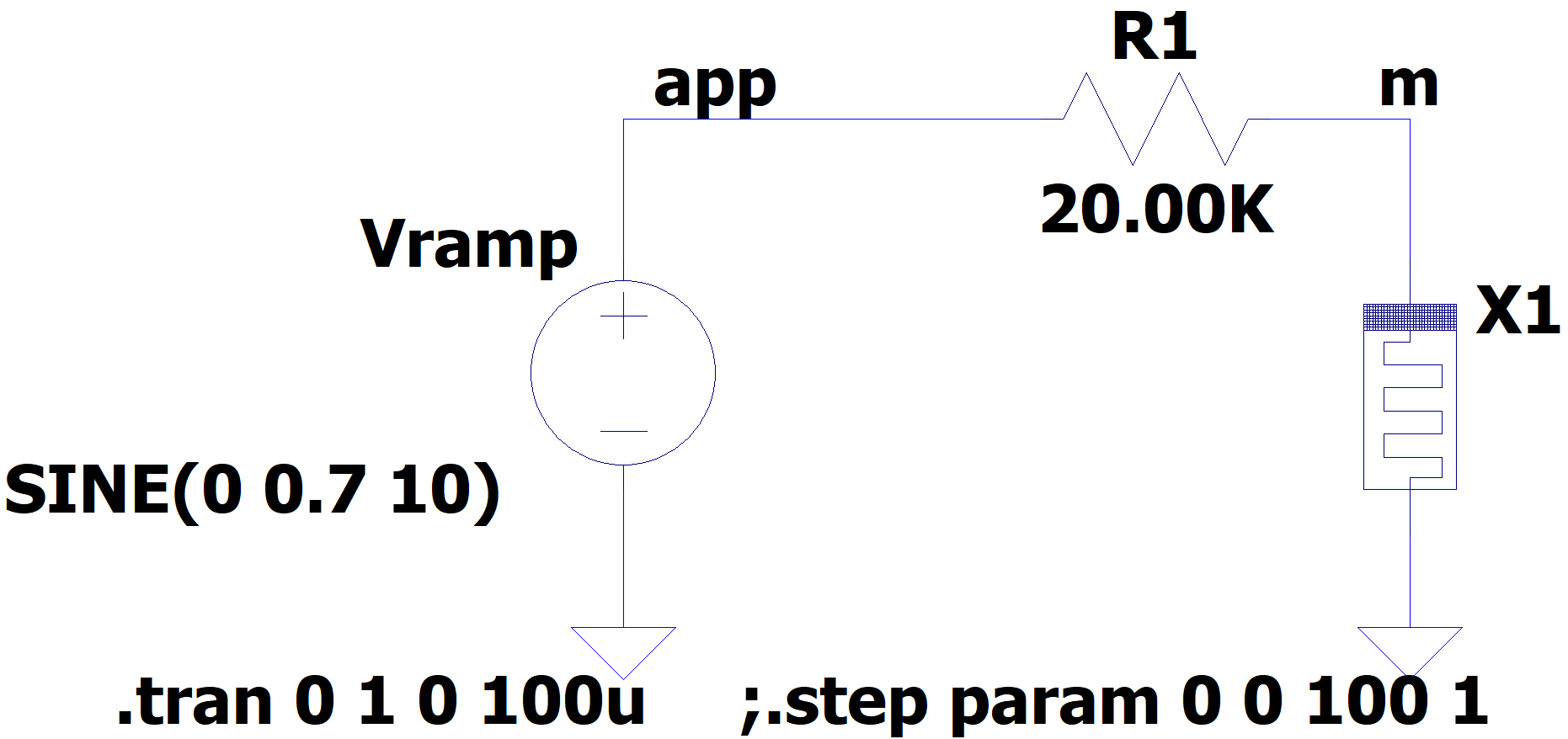}
	\caption{Physical Model with variability}
\end{figure}

\begin{figure}[ht!]
	\centering
	\includegraphics[width =0.7\textwidth]{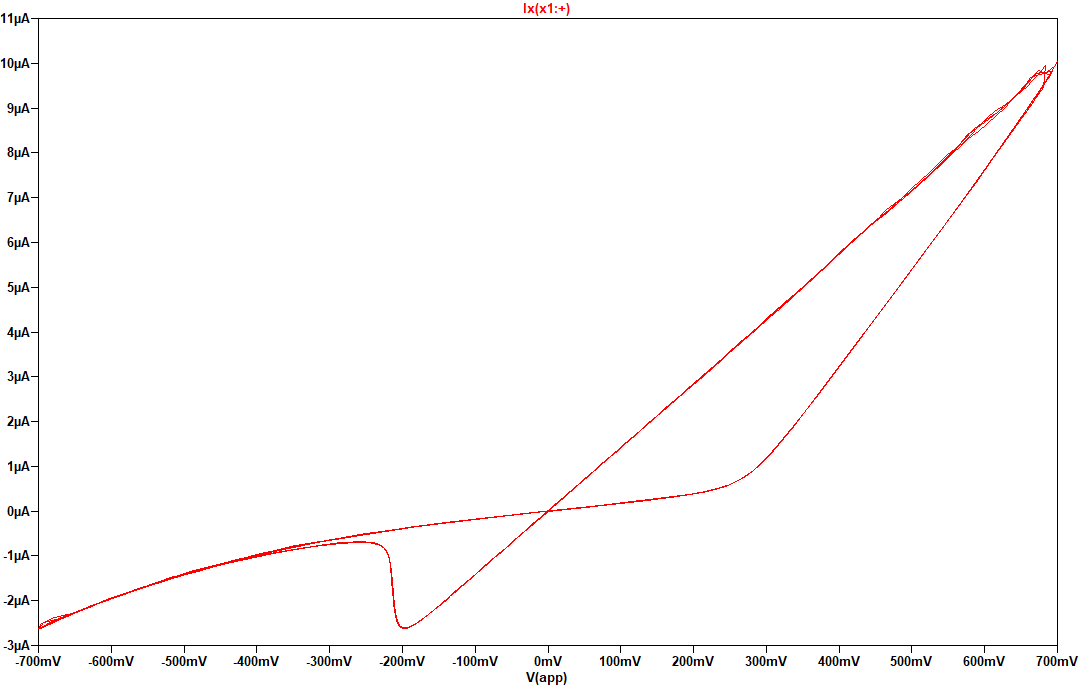}
	\caption{LTspice simulation with variability}
 \label{LTspice simulation with variability}
\end{figure}
The data used as a starting point to create the model are from paper \cite{Aguirre202214}, which are shown in Table \ref{Init variability params}.

\begin{table}[h!]
\centering
\begin{tabular}{|l|l|l|l|}
 \hline
 roff = 10$\Omega$ & ri = 50$\Omega$ & ron = 10$\Omega$ & ion0 = 4.95µA\\
 \hline
 ioff0 = 2.48µA & aon = 1.71 & aoff = 2.58 & H0 = 0 \\
 \hline
etas = 50 & etar = 250 & Vs = 0.2V & Vr0 = -0.02V \\
 \hline
Vt = 0.05V & isb = 7µA & gam = 0.35 & sigion =  0.1\\
 \hline
sigioff = 0.25 & sigvr = 0.01 & sigsb = 0.02 & RPP = 1E10\\
 \hline
 
\end{tabular}
\caption{Init variability params}
\label{Init variability params}
\end{table}

To utilize this model created in LTspice, it was decided to integrate it with Python in order to streamline parameter changes and, in the future, facilitate the development of genetic algorithms and neural network creation. For this purpose, the PyLTSpice Python library was used, which enables a quick and straightforward connection with the models created in LTspice. Additionally, this library allows running the model with desired parameters directly from Python. Thanks to these features, parameter adjustments were made by creating a genetic algorithm.

\subsection{Genetic Algorithm}
\label{sec:genetic}

A genetic algorithm was implemented to find the best parameters for the physical model. 

A genetic algorithm is an optimization algorithm that mimics the process of natural selection. It works by iteratively creating a population of solutions, evaluating their fitness, and then using genetic operators to create a new population of solutions. The genetic operators are used to introduce variation into the population, which helps to ensure that the new population can contain solutions that are better than the old population. The process is repeated until a satisfactory solution is found.

The algorithm was implemented in python, and the following steps were taken:
\begin{enumerate}
\item In each generation, the learning rate was decreased incrementally.
\item The current in $Vramp$ and the voltage in $app$ were obtained during the execution in LTspice with the current population, obtaining the simulated I-V curves.
\item The loss of each individual was calculated by mapping the simulation curve onto the real ones and calculating the square differences between them.
\item The average of losses was updated, and the population was sorted from the lowest to the highest loss.
\item The best individual was updated, and the half with the worst loss was eliminated while the remaining half was mutated.
\end{enumerate}

The algorithm was repeated until a satisfactory solution was found.

The following image shows the results obtained, being the blue graph the original data and the red one the data obtained with the model after the execution of the genetic algorithm.

\begin{figure}[h!]
	\centering
	\includegraphics[width =0.8\textwidth]{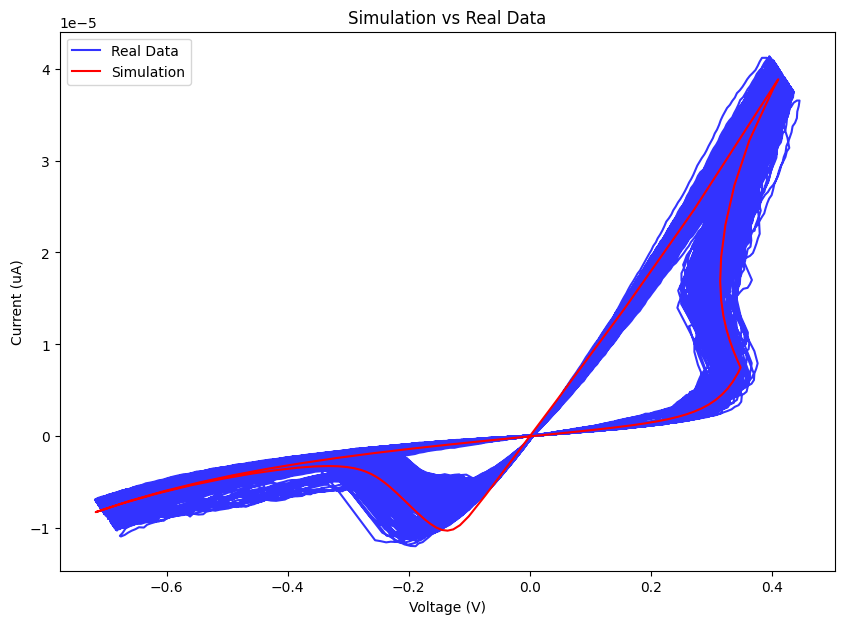}
	\caption{Genetic Algorithm results vs empirical curves}
\end{figure}

Next, this genetic algorithm was modified to incorporate the variability present in commercial memristors such as the one used by KNOWM. Memristors have the ability to modify their resistance based on the electric charge passing through them. Despite the stochastic nature of these devices, there are variations between different cycles or among different devices under the same conditions, thus demonstrating the aforementioned variability.

This new genetic algorithm was applied to all the parameters of our current parameters, resulting in the following values, as shown in the figure, with a good agreement between the experimental and simulated curves.
\begin{table}[h!]
\centering
\begin{tabular}{|l|l|l|l|}
 \hline
 roff = 10$\Omega$  & ri = 50$\Omega$ & ron = 10$\Omega$ & ion0 = 50µA\\
 \hline
 ioff0 = 2.5µA & aon = 1.72 & aoff = 2.7 & H0 = 0 \\
 \hline
etas = 17 & etar = 70 & Vs = 0.14V & Vr0 = -0.03V \\
 \hline
Vt = 0.1V & isb = 6.2µA & gam = 0.29 & sigion =  0.05\\
 \hline
sigioff = 0.07 & sigvr = 0.01 & sigsb = 2.04e-07 & RPP = 1E10\\
 \hline
\end{tabular}
\caption{Genetic variability params}
\label{Genetic data}
\end{table}

The 'steps' in each LTspice simulation could not be parallelized because all the parallel simulated steps used the same seed. As a result, the same values were obtained for the sigion, sigioff, sigvr, and sigisb parameters responsible for emulating the spoken variability. This meant that for each individual, all the steps performed used the same Gaussian distribution, resulting in no variability.

However, the execution of different individuals in the population could be parallelized, greatly speeding up the simulations. This was done from Python by launching multiple instances of LTspice, one for each individual. We were able to achive a speed up of up to $6.25X$.

\begin{figure}[h!]
	\centering
	\includegraphics[width =0.8\textwidth]{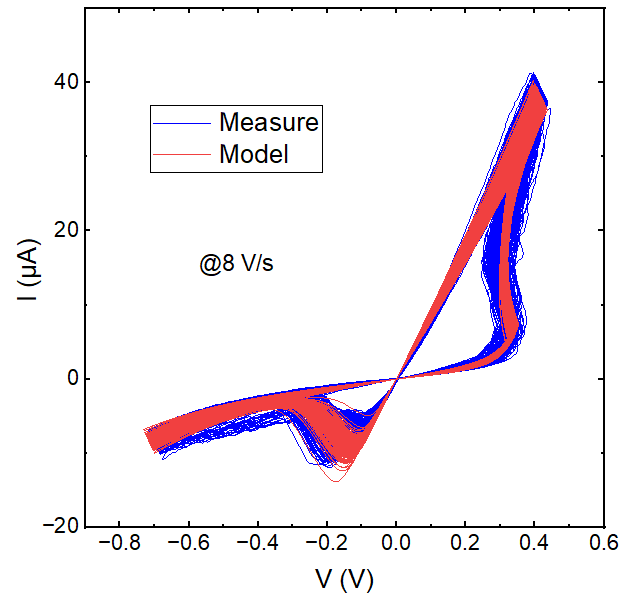}
	\caption{Genetic Algorithm with variability}
 \label{Genetic Algorithm with variability}
\end{figure}

\subsection{Results}

After using the genetic algorithm with variability, we obtain the final data for our model. The data shown in Table \ref{Genetic data} allow us to perform a simulation very similar to the hysteresis curves obtained during the experimental phase with the commercial KNOWM memristor.

\begin{figure}[h!]
	\centering
	\includegraphics[width =\textwidth]{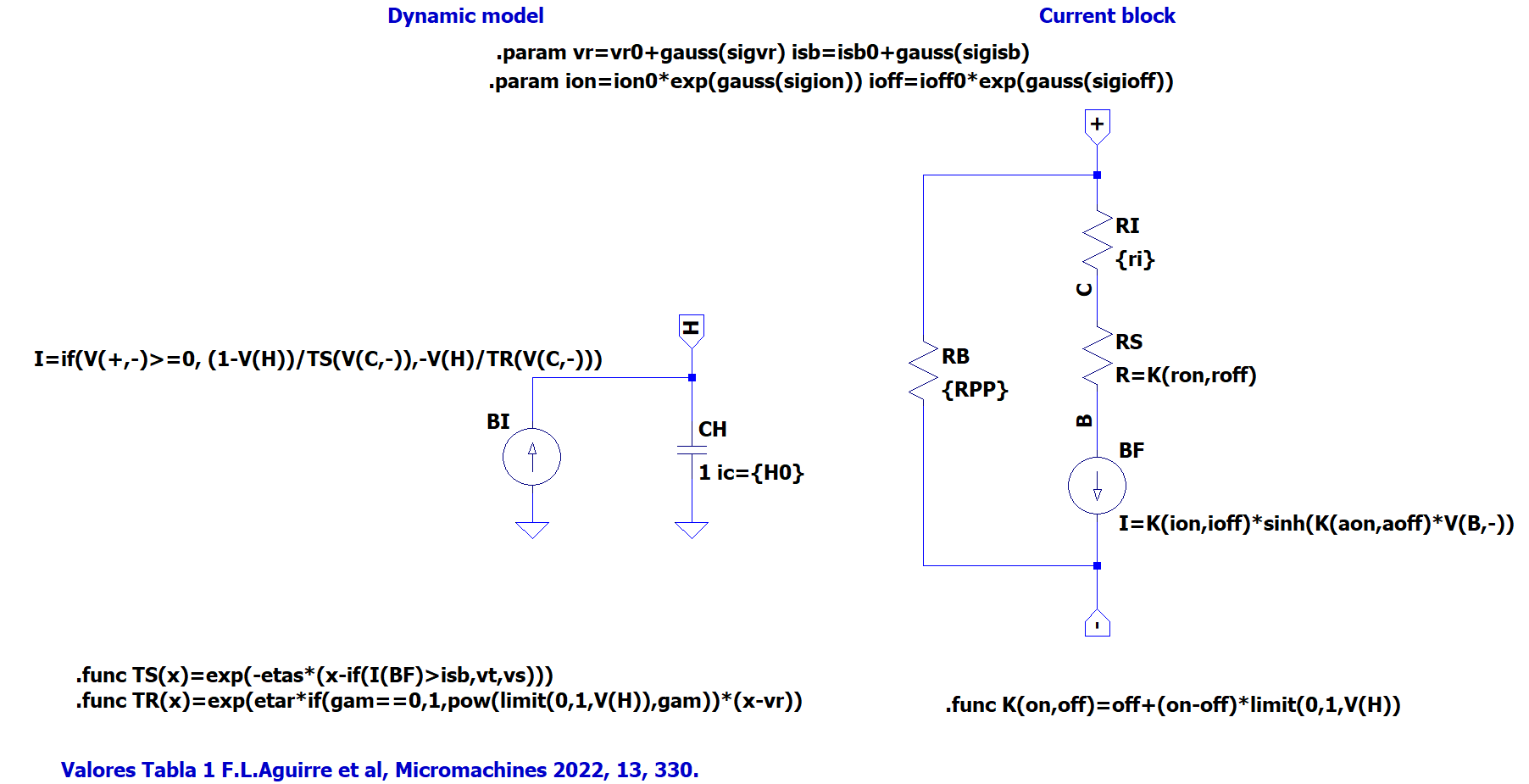}
	\caption{Memristor Physical Model with variability}
 \label{Memristor LTspice}
\end{figure}

The model, using these parameters, is finally represented as shown in Figure \ref{Memristor LTspice}.

\begin{figure}[h!]
	\centering
	\includegraphics[width =0.7\textwidth]{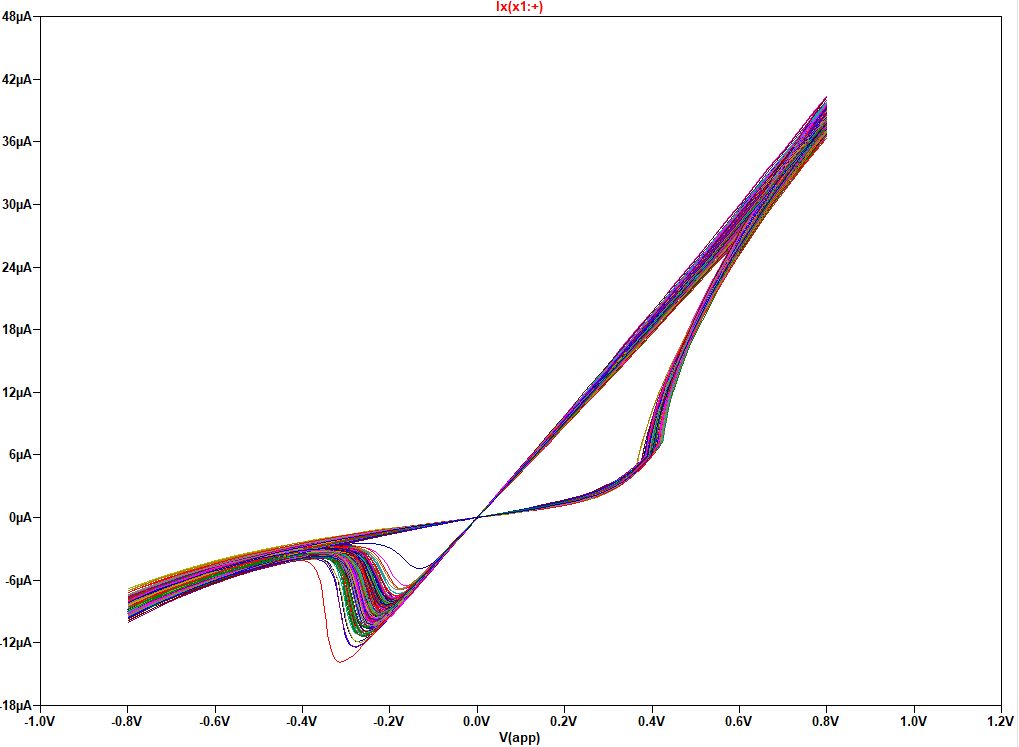}
	\caption{Memristor Physical Model Hysteresis}
 \label{Memristor LTspice Hysteresis}
\end{figure}

Using the best parameters in this new model, we obtain the following curves shown in Figure \ref{Memristor LTspice Hysteresis} with 200 steps and variability. As can be seen, these curves accurately simulate the behavior of a real memristor. These data are shown in Figure \ref{Genetic Algorithm with variability}, directly comparing them with the experimental curves obtained with the KNOWM memristor. The blue curves represent the actual measurements, while the red curves represent the model.

As mentioned before, due to the limited availability of physical memristors to simulate a complex neural network like MNIST, it is necessary to use a physical model that accurately emulates the behavior of a memristor. With the obtained data, it is considered that the resulting model is faithful enough to the commercial memristor used to evaluate these devices as a viable alternative in neural network creation.

\section{Neural Network Implementation}

We now have a suitable physical model for our memristors and the parameters that best fit our empirical data in simulations. Moving forward, we will no longer be constrained by the limited number of memristors on the board and their configuration.

We can start to design and test circuits by means of hardware simulation in LTSpice, using as many memristors as our processing power will allow and in whichever configuration we desire.

\subsection{Inference in Neural Networks}
\label{sec:inference}
We will design a circuit to perform neural network inference by means of analog computations. Before explaining the circuit design, we will do a brief explanation on how an inference step is calculated form a mathematical point of view.

\subsubsection{Fully Connected Neural Network}

\begin{figure}[h!]
	\centering
	\includegraphics[width =0.7\textwidth]{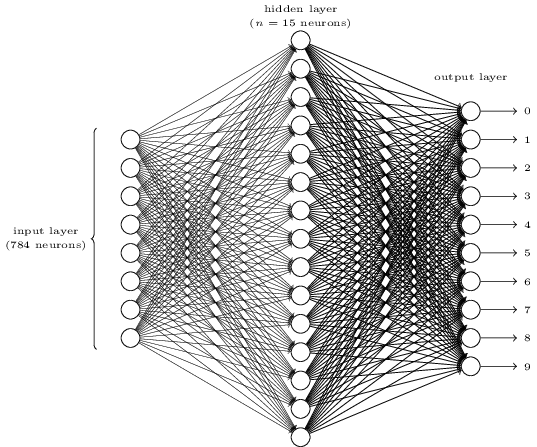}
	\caption{Neural Network illustration from \cite{neuralNetworks}}
\end{figure}

A neural network is made up of an ordered sequence of layers, which are in turn composed of an array of neurons. Each layer of neurons is activated, one after the other, depending on the values of the output of the immediately previous layer. In a fully connected neural network, each and every one of it's layers is fully connected to the next one, meaning all neurons receive as inputs the outputs of every other neuron in the preceding layer. Each neuron activation value is calculated as follows: 
\begin{center}
$z =\sum_j w_j x_j + b$
\end{center}
Where $x_j$ is the value of the output of the $j^{th}$ neuron in the previous layer, $w_j$ is the neuron's weight associated with the $j^{th}$ output and $b$ is the bias parameter of the neuron. If we have $w$ and $x$ as vectors, we need simply to calculate their dot product and then add $b$.

Afterwards, an activation function, such as sigmoid or relu, is applied to $z$ to produce the final output that will be fed as an input to the next layer.

The calculations for each layer can be easily made in parallel, since the activation of one neuron doesn't depend in any way on the activation of other neurons within the same layer. This way, instead of calculating dot products for each neuron, we perform a matrix multiplication as follows:
\begin{center}
$z =W_l * A_{l-1} + B_l$
\end{center}
Where $W_l$ is the matrix containing the weights of the neurons of layer $l$. Each row corresponds to a different neuron, and each column to the associated neuron of the previous layer. $A_{l-1}$ is the vector of outputs from the previous layer, or the inputs if $l$ is the first layer. Lastly, $B_l$ is the vector of biases of the layer $l$ each of them corresponds to a neuron.

The inference process is the calculation of the output of given an input. To do this, we only need to follow these steps for each layer in order, starting with $l=1$:
\begin{enumerate}
  \item Calculate $z = W_l * A_{l-1} + B_l$.
  \item Apply the activation function to $z$ to get $A_l$.
  \item If $l$ isn't the last layer, do $l+1$.
\end{enumerate}

\subsubsection{Convolutional Neural Network}
The other type of neural networks that we will implement are convolutional neural networks. These types of networks usually consist of a set of convolutional layers followed by some fully connected layers. They are widely used for image classification tasks, among many other things, because they retain the ability to generalize at a higher training efficiency.

The inference in these last fully connected layers works in the same way as previously described, so we will only focus on the convolutional layers in this section.

\begin{figure}[h!]
	\centering
	\includegraphics[width =0.7\textwidth]{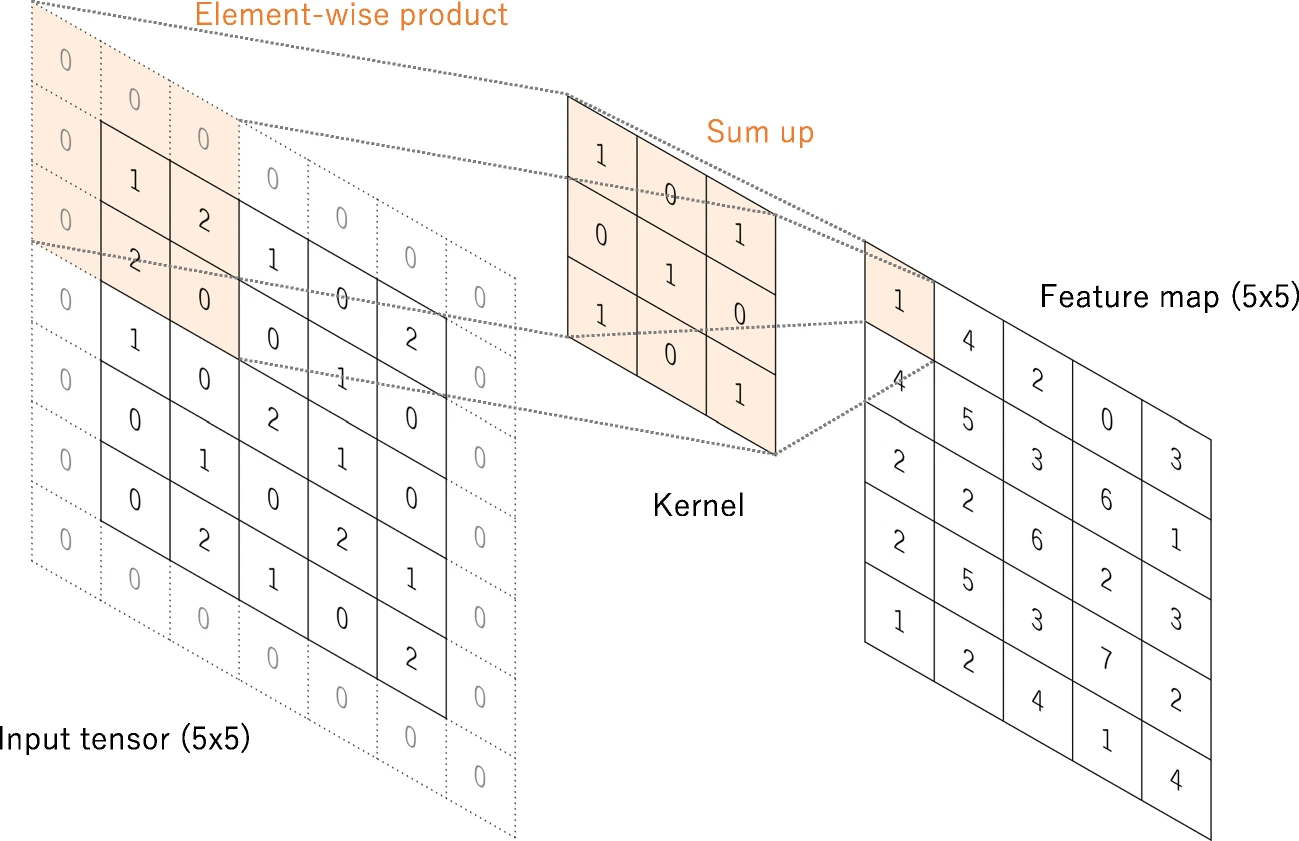}
	\caption{Convolutional layer illustration from \cite{Yamashita2018}}
\end{figure}

Convolutional layers use kernels, which are square matrices of a given size, typically 3x3 or 5x5. The activation value of a neuron is calculated only using a kernel sized window of neurons from the previous layer. The weights of the kernel are common to all neurons, which simplifies the structure. It is often explained as a sliding window, that multiplies with the input image element-wise, adds up all the terms and adds a bias.

Usually, more than one kernel is used, generating outputs with more than one channel, i.e., with more than one value for each position in the resulting feature map. This way, we have a third dimension, the channels. There will be as many kernels as $Channels_{in} * Channels_{out}$, meaning that for every channel of our feature map, there will be one kernel for each channel in the input; their resulting values will be added up.

\begin{figure}[h!]
	\centering
	\includegraphics[width =0.7\textwidth]{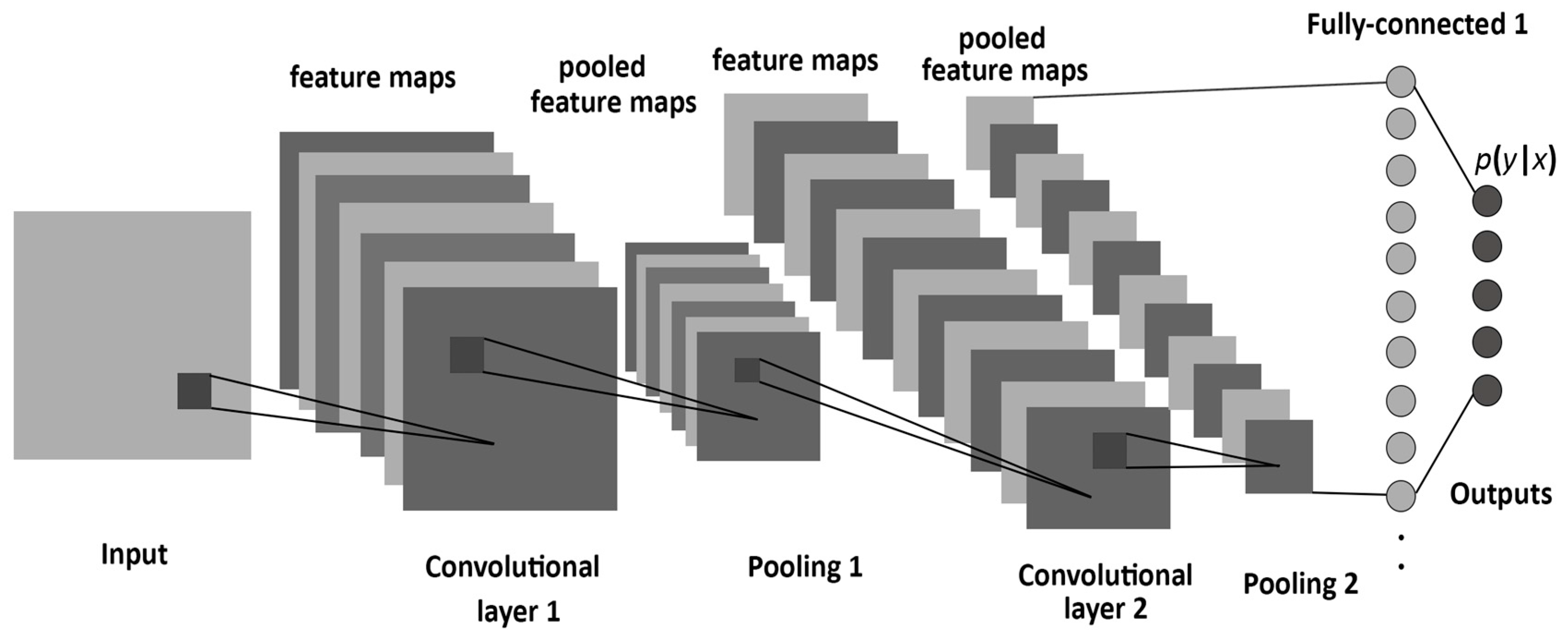}
	\caption{Convolutional network illustration from \cite{e19060242}}
\end{figure}

Unless some type of padding is used, the size of the output feature map will be smaller than that of the input. The reason is that values are not calculated for positions at the edges where the kernel window does not entirely fit. 

To further reduce the size of the feature maps, techniques such as pooling or strides can be used. A pooling layer divides the feature map into windows, usually $2x2$, and reduces them to a single value, either by taking the maximum value of the the window or by averaging them all. Strides allow us to achieve a similar result with a very small hit to quality, less complexity and less computations. The idea is that, instead of applying the kernels to every single possible position, we skip some in between.

Taking all of this into account, the resulting feature map will be a tensor made up of as much matrices as output channels and each of the matrices will have a size equal to the size input matrix minus the lost edges divided by the stride.

A forward pass in a convolutional layer is as follows:
\begin{algorithm}
\caption{Convolutional layer}
\label{conv_algorithm}
  \begin{algorithmic}[1]
  \State $k\gets size_{kernel}\div 2$\;
  \State $(size_{fmap}\gets size_{input} - k*2)\div stride$\;\Comment{Size of the resulting feature maps}
  \For{\texttt{$cout$ in $Channels_{out}$}}
        \For{\texttt{$i$ in $size_{fmap}$}}
                \For{\texttt{$j$ in $size_{fmap}$}}
                        \For{\texttt{$cin$ in $Channels_{in}$}}
                            \For{\texttt{$ki$ in $[-k$, $k]$}}
                                \For{\texttt{$kj$ in $[-k$, $k]$}}
                                    $H[cout][i][j] \gets H[cout][i][j] + input[Cin,i*stride+ki+k,j*stride+kj+k] * kernel[Cout][Cin][ki+k][kj+k]$\;
                                \EndFor
                            \EndFor
                        \EndFor
                    \State $H[cout][i][j] \gets H[cout][i][j] + B1[cout]$\;\Comment{Adding the biases}
                \EndFor
        \EndFor
  \EndFor
  \end{algorithmic}
\end{algorithm}

\subsection{Hardware Design}
\begin{figure}[h!]
	\centering
	\includegraphics[width =0.7\textwidth]{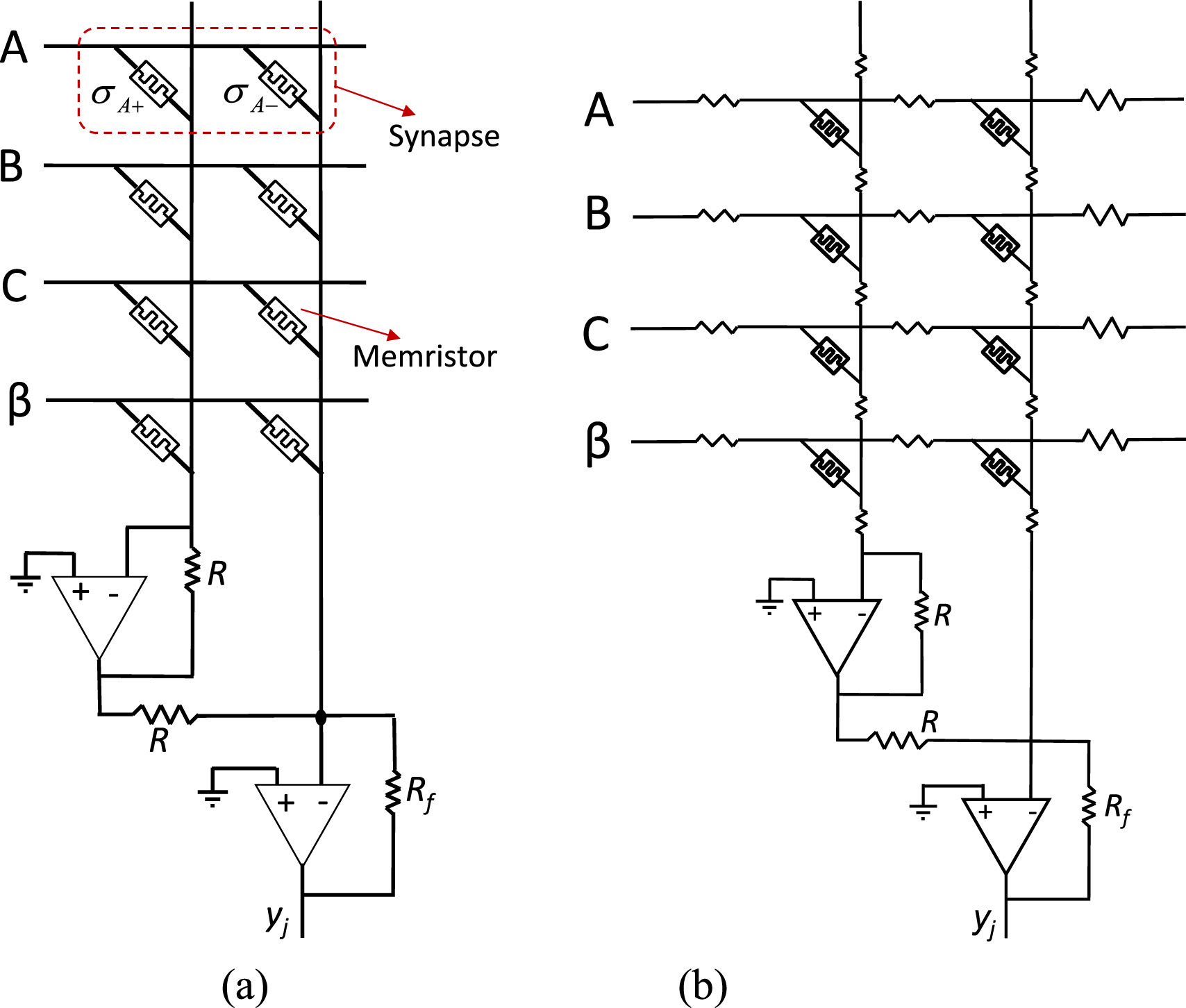}
	\caption{Memristor-based neuron circuit. A, B, C are the inputs and yj is the output. Figure from \cite{HASAN201731}}
 \label{fig:cross_bar}
\end{figure}

Once we know what operations we will need to implement in our circuit, we can start with the design. We will use a crossbar structure with synapses of two memristors, similar to the one used in \cite{HASAN201731} and which can be seen in figure \ref{fig:cross_bar}.

This structure is very widely used in analog computation, as it allows for very simple matrix multiplication. The inputs are the voltages in nodes $A$, $B$ and $C$. By Ohm's law, the resulting current will be $V/R$ or, using the conductance: $I=V*G$. All the currents that carry the value of an input multiplied by the conductance of the corresponding memristor are added up by virtue of being connected to the same cable.

\subsubsection{Design of a Fully Connected Neural Network}
\label{subsection:design}

Going back to the preceding explanation of inference, our inputs and activations of the previous layer are represented as the magnitude of the input voltages, which can be both positive or negative. Our weights are represented by the value of the conductance of the pair of memristors that form each synapse. 

A neuron is a pair of columns connected to an operational amplifier. Its weights are its synapses, placed in the row corresponding to their associated input. In figure \ref{fig:cross_bar} only one neuron is depicted, adding more is as simple as replicating the structure, attaching more pairs of columns to the left.

We use the synapses to be able to represent negative weights: if the conductance of the first memristor $G1$ is higher than the one of the second $G2$, it represents a positive weight of value $G1 - G2$, otherwise, the weight will be negative. To achieve this, we need a sub-circuit that receives the currents from both columns and subtracts the value of second from the first one. In turn, this device, known as an operational amplifier, will give us the result in terms of a voltage, ready to be fed to the next layer of neurons or to be read as the final output. The magnitude of this voltage will be the resulting current multiplied by a resistance $R_f$.

The bias term of each neuron can be treated as a weight, meaning its value is determined by the conductance of a synapse. But, instead of receiving a variable input, it is always supplied a constant amount of voltage, making it independent of the input. Its addition to the dot products is carried out by being connected to the same cable.

Putting all this together, the output of a neuron, pending the activation function, will be the following:

\begin{center}
$z = R_f*([\sum_j V_j*G1_j] + \beta*GB1 - ([\sum_j V_j*G2_j] + \beta*GB2))$

or

$z = R_f*([\sum_j V_j*(G1_j-G2_j)] + \beta*(GB1-GB2)$
\end{center}

Where $\beta$ is a constant, $GB1$ and $GB2$ are the conductances of the memristors that make up the synapse that represents the bias term. The same for each pair of $G1_j$ and $G2_j$, which are the weights. 

We now have a way of calculating $z = W_l * A_{l-1} + B_l$. It's worth noting that the cross-bar structure doesn't perfectly map to the matrix  $W_l$, since the weights of a neuron were the rows of $W_l$ but are now the columns of the cross-bar. We are only missing a way of applying the activation function. There are electrical circuits than can effectively approximate these functions, but since it isn't necessarily relevant for our purpose, we chose to simulate then behaviorally instead of designing the circuits for them.

\begin{figure}[h!]
	\centering
	\includegraphics[width =0.7\textwidth]{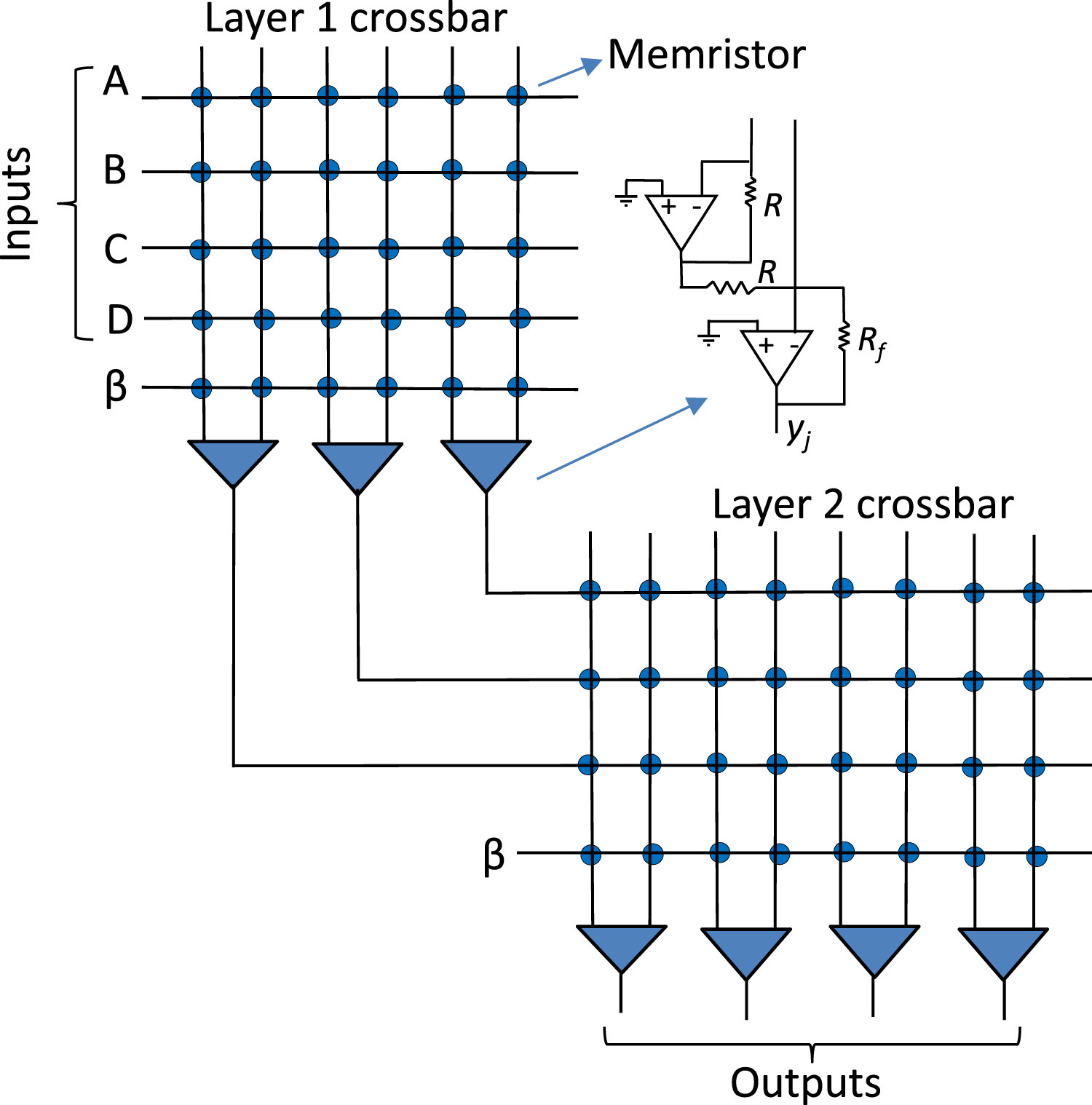}
	\caption{Multi-layer neural network. Figure from \cite{HASAN201731}}
 \label{fig:multi_layer}
\end{figure}

To have a complete circuit for a neural network with more than one layer, we simply concatenate another cross-bar structure that takes as inputs the voltage outputs of the last one, as in figure \ref{fig:multi_layer}.

\subsubsection{Design of a Convolutional Neural Network}
Looking at algorithm \ref{conv_algorithm} one would be correct in assuming that its circuit design won't be as straight forward as that of the fully connected network. There is hardly an easy way of visualizing it.

Regardless of that, the calculation can be parallelized not very differently than how we did for the fully connected layers. We won't need to use any other sub-circuit or technique than the ones we already have, since we will only need to do addition and multiplication and we can use the same operational amplifier as before.

The difficulty of implementing this circuit lies in its interconnections. However, as we will explain in the following section, we are not designing the circuits manually, but rather using scripts, which will make this task feasible.

Some noteworthy differences are in the kernels, which contain the synapses representing the weights. Whereas in the fully connected layers each weight could be unique, the kernels and its synapses will need to be replicated for as many times as output values the output feature map contains. This is not a problem for us, since we are only interested in performing inference, but it is an important consideration for training.

\subsubsection{Time and Power Consumption Estimations}
We will now estimate the time of inference and the energy usage of a fully connected neural network with a hidden layer of 20 neurons.

For the time we are going to do a first order estimation treating the columns that represent each neuron as if they were in series with respect to the input. In reality they are neither in series nor in parallel, which complicates this analysis. The total time of the circuit will be determined by the result of $T = R*C$ where $R$ is the total resistance of the interconnection cables, which we will assume to be $5\Omega$, and $C$ is the total capacitance. To calculate the total capacitance, we must add the capacitance of all the individual memristors that are in parallel.

To do this, we have to look back at the circuit design. Since we are treating each neuron as parallel to each other, and they all have the same number of memristors, we only need to take in to account the total number of memristors for a single neuron. This will be the same as weights the neuron has, which in turn, corresponds to as many inputs as it processes. Each weight is composed of two memristors, but they are in series. Therefore, the amount of memristors in parallel in a given layer will be equal to the number of inputs it has. For a $16x16$ image, we have $256$ memristors in parallel.

Now, the second layer will receive $20$ inputs from the first layer, which means that it will have $20$ memristors in parallel. Adding this up, we have a total of $276$ of them in parallel. If we assume a typical capacitance of $20pF$ for each memristor, then $T=5*276*20pF$ which gives us a result of $27,6ns$, or a frequency of $36,2MHz$.

For the power consumption, we measure the current in the operational amplifier, which has a value of $0.000754A$ and we multiply it by the voltage supplied $5V$, which gives a result of $3.77mW$ for each neuron. Since in this example we have $20$ hidden neurons plus $10$ output neurons, the total power consumption is $0.1131W$

\subsection{LTSpice Implementation}
The size and complexity of the circuits for the neural networks make it impractical if not impossible to draw. Instead, we will use Netlists, which allow us to describe a circuit and its interconnections using plain text and can be used to compile circuits and do simulations in LTSpice.

In the Netlist, aside from the connectivity description, we can add simple components such as voltage sources or resistors, but for the rest, we will need to call sub-circuits. We only need two of them: one for the memristors and another one for the operational amplifiers. 

We already have the circuit for the memristor, as seen in chapter \ref{chap:physical_model} and we can easily extract a Netlist from it. We chose to use it coupled with the $10K\Omega$ resistor with which our real memristors came with, since that is the configuration we have characterized. It receives as a parameter the value of H0 so that we can give it an initial state, i.e., set a given conductance.

\begin{figure}[h!]
	\centering
	\includegraphics[width =0.9\textwidth]{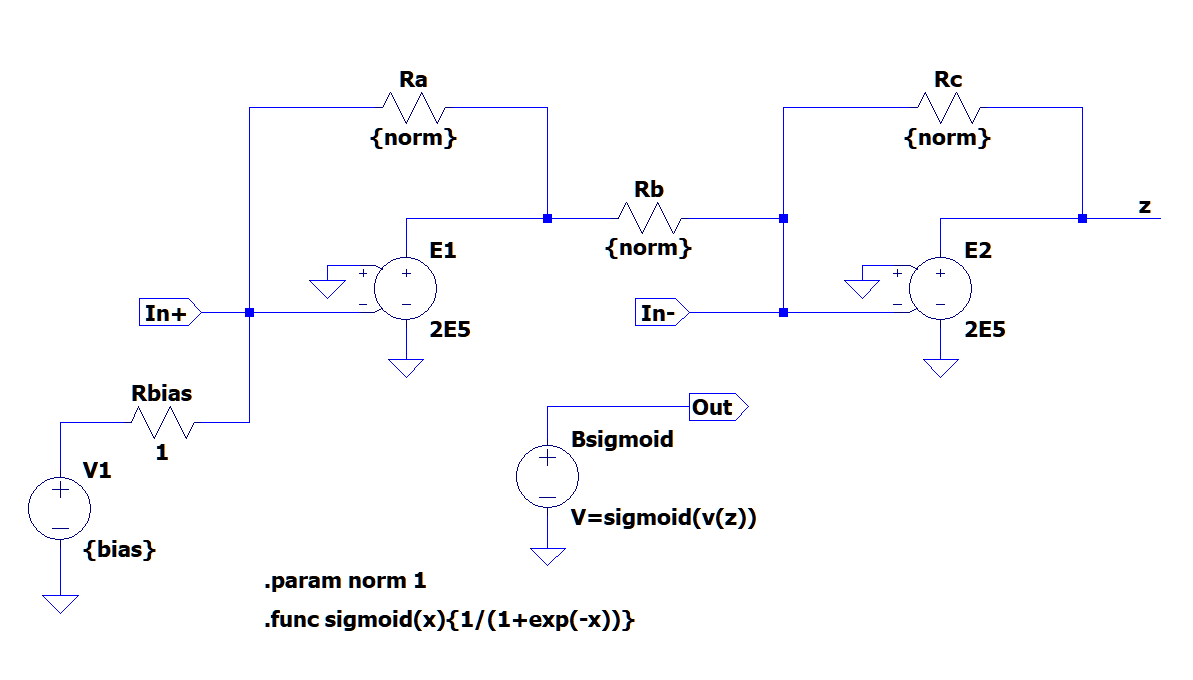}
	\caption{Operational Amplifier}
 \label{fig:op_amp}
\end{figure}

For the operational amplifier, we built the circuit shown in figure \ref{fig:op_amp}, from which we extracted its Netlist. We also apply the activation function in here, reason why we have slight variations of this sub-circuit, differing only in this respect.

As an example of how a Netlist describes a circuit, the following text is equivalent to figure \ref{fig:op_amp}:

\qquad \textit{.subckt neuron In+ In- Out PARAMs: norm=1 bias=0}

\qquad \textit{E1 N001 0 0 In+ 2E5}

\qquad \textit{Ra N001 In+ {norm}}

\qquad \textit{E2 z 0 0 In- 2E5}

\qquad \textit{Rc z In- {norm}}

\qquad \textit{Rb N001 In- {norm}}

\qquad \textit{Bsigmoid Out 0 V=sigmoid(v(z))}

\qquad \textit{Rbias N002 In+ 1}

\qquad \textit{V1 N002 0 {bias}}

\qquad \textit{.func sigmoid(x){1/(1+exp(-x))}}

\qquad \textit{.backanno}

\qquad \textit{.ends}

It receives as parameters the value of bias, which we never used since we manage them as described in subsection \ref{subsection:design}, and the $norm$ parameter, which is the value that will multiply the result of the addition; $R_f$ in subsection \ref{subsection:design}.

To write the Netlist for the circuit of each neural network, we wrote script in Python that receives the following information as parameters:
\begin{enumerate}
    \item The name of the file to write on.
    \item The size of the image.
    \item The amount of input channels (1 for B\&W, 3 for RGB).
    \item Number of output neurons.
    \item The weights and biases of each layer, adapted to their corresponding H0 equivalents, see section \ref{sect:param_trans}.
\end{enumerate}
With this information, the script calls functions to write each of the layers, depending on the type of network. We successfully implemented functions to be able to do a linear layer form both a preceding convolutional layer or another linear layer, a convolutional layer, and the operational amplifiers for both types of layers.

The functions that create the linear and convolutional layers aren't any more complicated than following the steps and algorithms detailed in section \ref{sec:inference}. Instead of having variables where we add up the products of our inputs and weights, we now simply use the same output cable where the currents add up. For example, a synapse representing a negative weight will be written like this:

\qquad \textit{Xwh39+	H627	H739+	memristor PARAMS: Hvalue=0}

\qquad \textit{Xwh39-	H627	H739-	memristor PARAMS: Hvalue=1.959469e-06}

Where $Xwh39+$ and $Xwh39-$ are the names of the two memristors that make up a synapse; $H627$ is the input cable for both. In the output of the first memristor, node $H739+$ we will get the current that results when multiplying the voltage in the input with the conductance of the memristor with $H0=0$, which will be very close to $0$. The same goes for $H739-$, except this time we will have some current. For a positive synapse, the $H0$ of the second memristor would be $0$ and the first one would be greater than $0$; and if the weight happened to be $0$, then they would both be $0$.

For the biases, the input cable is always $Bias$, which has a fixed voltage.

The operational amplifiers work in a similar way, we only need to connect them to the corresponding inputs and outputs:

\qquad \textit{XNeuron3	H5+	H5-	OUTPUT0	simple PARAMS: norm=1000 bias=0}

Where $XNeuron3$ is the name, $H5+$ and $H5-$ are the two inputs carrying the currents that will be subtracted and the result will be multiplied by $norm=1000$ and converted to voltage. The activation function in this case would be the identity since the type is $simple$.

With these building blocks, we can very easily implement different architectures that can differ in input channels, size of images, size of kernels, weights and biases, number and type of layers, output neurons and activation functions.

The Netlists we built for the circuits are, in fact, sub-circuits themselves. In section \ref{sec:simulation} we will explain the reasons, which have to do with the management of the inputs for the simulations.

\subsection{Parameter Convertion}
\label{sect:param_trans}
We now have the circuit designs and the means to build Netlists for them. We need only to acquire the appropriate weights and biases to be able to start the simulations.

Doing a short recapitulation, our memristor sub-circuits only take as an argument the $H0$, but we have defined the weights and biases of the memristor synapses as their conductivity, not their $H0$. Therefore, we need a way of making the conversion from conductivity to $H0$.

\begin{figure}[h!]
	\centering
	\includegraphics[width =0.8\textwidth]{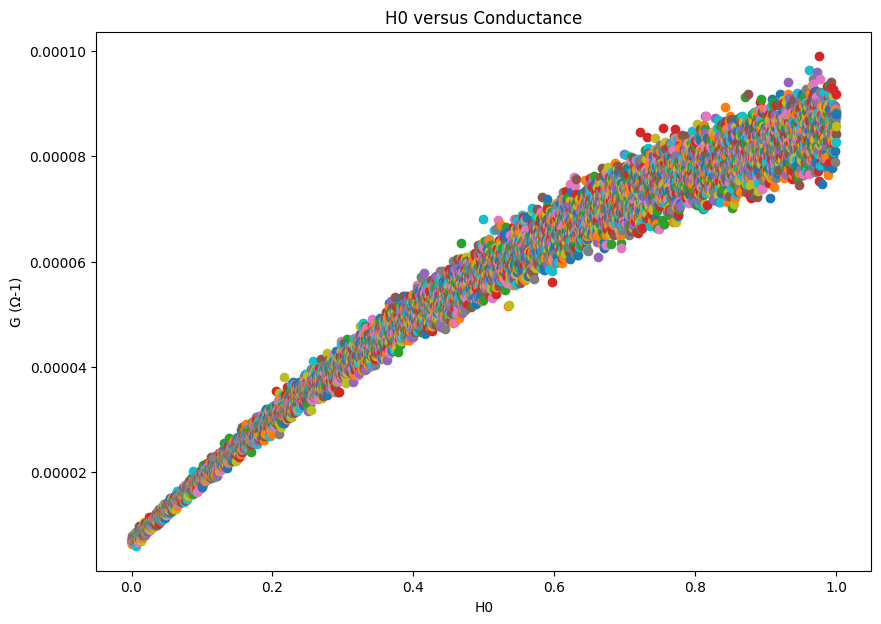}
	\caption{H0 (state variable of the device) versus the conductance}
 \label{fig:H0_G}
\end{figure}

The range of values that $H0$ can take is between $0$ and $1$. To see how they relate to the conductivity, we run an LTSpice simulation on our memristor model that steps through 10.000 values of $H0$. We measure the conductance of the device at each point, while under a reading voltage of $0.1V$. The results can be seen in figure \ref{fig:H0_G} where the variability is very noticeable.

\begin{figure}[h!]
	\centering
	\includegraphics[width =0.8\textwidth]{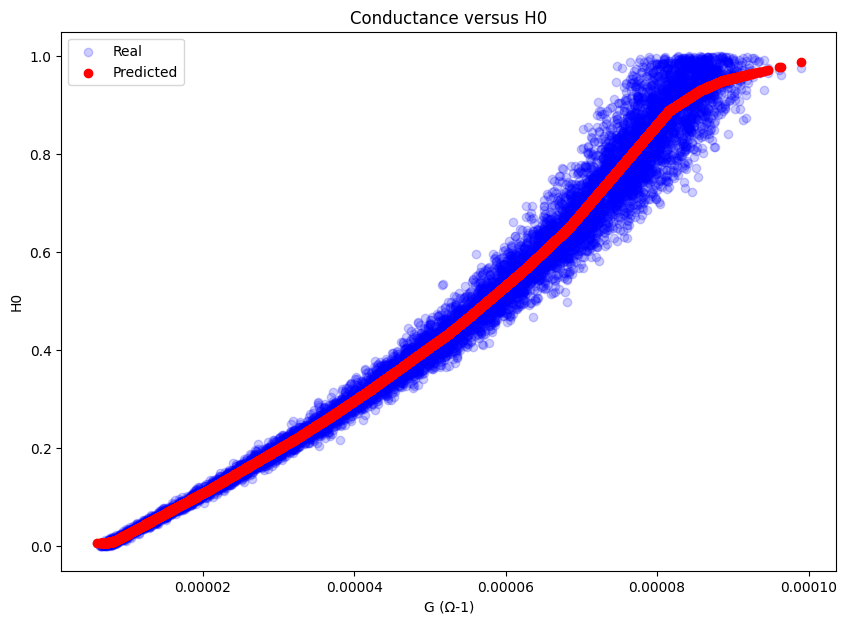}
	\caption{Results of the model trained to approximate H0 for a given conductance}
 \label{fig:G_H0}
\end{figure}

We chose to train a small neural network to approximate the value of $H0$ that should be set when trying to achieve a given conductance. The results are in figure \ref{fig:G_H0}. We will use this model to convert the weights we will obtain in the next section to be able to pass them as arguments to our Netlist builders.

From figure \ref{fig:H0_G} we have also obtained the effective range for our weights and biases, which can't be any greater than $1e-4$. We chose to place the limit at $8e-5$, since that seemed like a good cut off point that, regardless of the variability, was almost always reachable with $H0=1$. As we are using two memristor synapses and are able to represent negative values, our real range is $(-8e-5, 8e-5)$.

\subsection{Digital Training}
Now that we know the restrictions for our inputs, weights and biases, we can try and obtain the parameters for the simulations of our circuits. For this purpose, we will use the PyTorch framework.

As previously mentioned, we will use the MNIST data set for our experiments, since it is a widely recognized and computationally light problem. Furthermore, we will be resizing the images form their $28x28$ original pixel size to $16x16$ and $12x12$. The reason behind this decision is the fact that the time it takes to simulate an inference step exhibits a quadratic growth in relation to the size of the input. We will use the high-quality LANCZOS interpolation method to reduce the quality penalty of the resize as much as possible.

\begin{figure}[h!]
	\centering
	\includegraphics[width =0.5\textwidth]{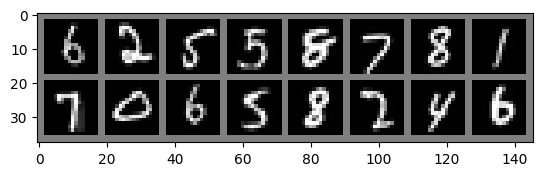}
	\caption{Scaled 16x16 MNIST examples}
 \label{fig:mnist}
\end{figure}

In the MNIST data set from the Torchvision package that we will be using, the images are in a gray scale with values from $0$ to $1$. These are the values that we will feed as inputs to our circuits, and, as we saw in section \ref{sec:char_results}, our memristors will change state under more than $0.25V$, which will alter the weights, impacting the performance of the network. To avoid this, we will apply a transformation to ensure that the inputs are contained within the safe interval from $-0.1V$ to $0.1V$.

The next step is to ensure the weights and biases are also constrained to the interval established in the previous section, $(-8e-5, 8e-5)$, since that is the available range for the conductance of our synapses. There is another important factor that further restricts the permissible interval of the biases, according to our physical design and as explained in subsection \ref{subsection:design}, the synapse that represents the bias is multiplied by a constant input voltage. Therefore, since that voltage has to be smaller than $0.25V$ to not introducing state changes, the real range of the biases will be further reduced. We chose $0.1V$ to simplify calculations, meaning our biases will have to be between $-8e-6$ and $8e-6$.

Since neural networks are not linear functions, we cannot simply train the network normally and then divide the value of the weights and biases to make them fit their intervals. This is because the activation functions are non-linear by design. Instead, we will attempt to train the networks directly with the restrictions.

This process required a great deal of trial and error, but we found a way of doing it without any loss in the performance of the network when compared to an equivalent one trained without restrictions. To do it, we had to follow these three steps:

\begin{enumerate}
    \item Initialize the weights and biases with a normal distribution with mean 0 and standard deviation of $2e-5$ for the weights and $2e-6$ for the biases.
    \item Use a weight and bias clipper after each gradient descent step.
    \item Avoid the vanishing of the activations layer to layer.
\end{enumerate}

\begin{figure}[h!]
	\centering
	\includegraphics[width =0.8\textwidth]{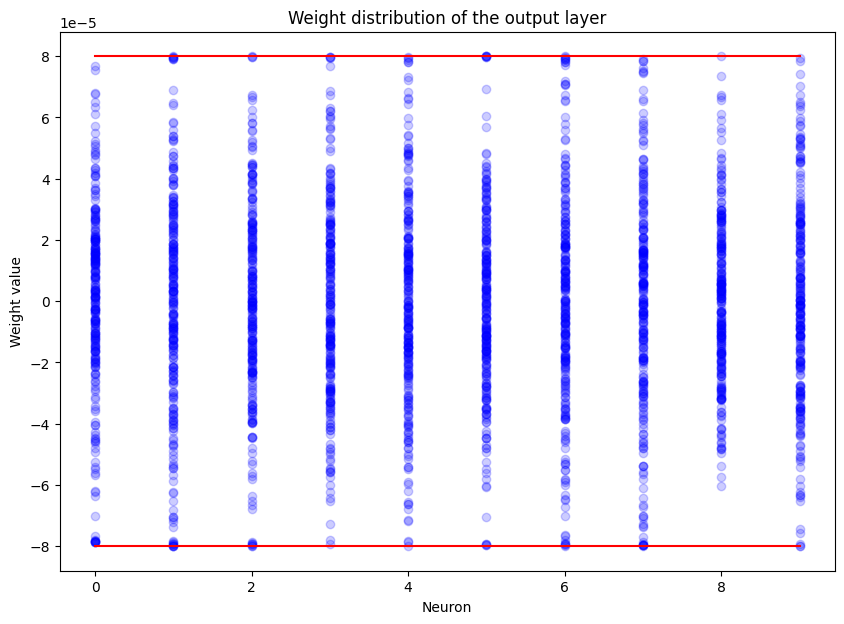}
	\caption{Weight distribution of the output layer of a neural network trained with parameter range restrictions}
 \label{fig:weight_distri}
\end{figure}

The first point is necessary so that the starting value of the parameters meet our restrictions. The second point is to ensure that the weights and biases stay within the valid range. To achieve this, we define a simple function that goes through all the parameters of the network and, if any of its weights is larger than $8e-5$, it sets its value to be exactly that and vice versa with weights smaller than $-8e-5$. It does the same with the biases, but with their maximum and minimum values. We call this function after every step of the gradient descent, i.e., every time the parameters get updated. We can see the results in figure \ref{fig:weight_distri}.

The third point is less straightforward, and it will depend on the activation function we use. Since our weights and biases are so small, every output of a layer, pending the activation function, is magnitudes smaller than its input was. For example, if we have a fully connected layer of $100$ neurons a single input, the output will be, in the best case scenario, where all the weights are the maximum of $8e-5$: 

\begin{center}
$input*100*8e-5=8e-3*input$  
\end{center}

In a more realistic scenario, we would have hundreds of inputs and they would also be both positive and negative, the same as the weights, which would further cancel each other out.

If we are using RELU as the activation function, the values will continue to get smaller and smaller with each layer, which causes a problem similar to the vanishing gradient problem, and the network struggles to learn anything. Sigmoid presents a different set of problems that we will tackle later. 

To solve this, we used inspiration from our physical design. In subsection \ref{subsection:design} we explained how the operational amplifier multiplied the resulting current by the value of a resistor $R_f$. We will use the same concept here, multiplying by a fixed value the output of every layer. We will then use the same value for $R_f$ in the Netlist builder to ensure that the voltages don't vanish either, so we will have to make sure that it does not produce voltages outside the interval $(-0.1V, 0.25V)$ or else it would change the state of the memristors in the following layer.

Finding the perfect value for this purpose is complicated, since we can have very different configurations of weights, biases and inputs. We found through experimentation that $1e4$ seemed to work well in most cases, particularly in fully connected layers.

We were also able to implement sigmoid activation functions, and it's worth explaining the challenges it presented. Firstly, the output of a sigmoid is in the interval $(0, 1)$ and $sigmoid(0) = 0.5$ which is well above our allowed values. For this reason, we need to change the sigmoid for a $sigmoid/10$, making the output $(0, 0.1)$. Another new problem with the sigmoid function is that, if the inputs are very close to $0$, it will behave like a linear function. To avoid this, we do something similar to what we did before; we multiply the outputs before the activation function by a constant. For sigmoids, $1e5$ worked the best.

Having found a way to successfully train neural networks that meet the restrictions imposed by our physical design, we trained the following configurations to later simulate and compare results:

\begin{enumerate}
    \item Simple fully connected network, no hidden layers, for input sizes $12x12$ and $16x16$.
    \item Fully connected network with a $20$ neuron hidden layer, for input sizes $12x12$ and $16x16$. Using RELU as the activation function.
    \item Simple convolutional network with a single convolutional layer, with kernel 3, stride 2, and a fully connected output layer, for input sizes $12x12$ and $16x16$. Using RELU as the activation function.
    \item Convolutional network with two convolutional layers, with kernel 3, stride 2, and a fully connected output layer, for input sizes $12x12$ and $16x16$. Using RELU as the activation function.
    \item Fully connected network with a $20$ neuron hidden layer, for input sizes $12x12$. Using Sigmoid as the activation function.
\end{enumerate}

\subsection{Simulation}
\label{sec:simulation}
Once we have the Netlist builders for our circuits, the parameters for the neural networks and a way of converting them to H0 values. We can start the simulations.

Our goal was to do inference on the first $1000$ images from the test set, both for the digital network and for the simulated circuit. For the same reasons explained in section \ref{sec:genetic}, we could not parallelize all simulations, one input at a time, because that way the variability parameters always take the same values. But we still wanted to parallelize as much as possible because the simulations take a considerable amount of time.

With $32GB$ of memory, we can do up to $10$ simulations at a time of the biggest circuits among the ones tested, which speeds up the total execution time by around a factor of $5X$.

To be able to have both the parallelization and the steps through the variability parameters of the physical model, we settled on doing $10$ threats, each of which will process $10$ different inputs using steps. This way, a batch size of $100$ images is divided into sets of $10$, which will be processed by different simulations. We do $10$ steps instead of $100$ because for some reason unknown to us, beyond $20$ or so steps, it starts to take considerably longer per step.

This is the reason why the Netlists of our circuits are sub-circuits, they need to receive the input as a parameter that will change with every step up to $10$ times.

The function that performs the simulation does the following:
\begin{enumerate}
    \item Calls the appropriate Netlist builder for the circuit, which in turn writes to the file specified the design of the neural network.
    \item Includes the sub-circuits needed, i.e, the memristor, the operational amplifiers and the circuit just built.
    \item Creates an instance of the neural network, giving it variables as the inputs.
    \item Steps through the input variables with the values of the different input images.
\end{enumerate}
An example of 4 would be the following:

\qquad \textit{.step param Vx list 1 2 3 4 5 6 7 8 9 10}

\qquad \textit{.param paramin000 table(Vx, 1, -0.1, 2, -0.1, 3, -0.1, 4, -0.1, 5, -0.1, 6, -0.1, 7, -0.1, 8, -0.1, 9, -0.1, 10, -0.1)} 

In this case, since it's a pixel on the corner of the image, it will always take the same value $-0.1$ which is black.

\section{Results and Variability}
We can now compare how our hardware implementations perform against their digital counterparts on inference precision. We include the results in accuracy in table \ref{tab:results}. Where the column $Ratio$ is the ratio of the simulation accuracy by the digital accuracy and the $type$ column follows the following name conventions:

\begin{itemize}
  \item FC Simple: a fully connected network without hidden layers.
  \item FC Double: a fully connected network with a 20 neuron hidden layer. Using RELU.
  \item CV Simple: one convolutional layer followed by the output layer. Kernel size $3$, stride $2$, output channels $3$. Using RELU.
  \item CV Double: two convolutional layers followed by the output layer. Kernel size $3$, stride $2$, output channels: first layer $3$, second layer $6$. Using RELU.
  \item FC Double Sigmoid: a fully connected network with a 20 neuron hidden layer. Using Sigmoid.
\end{itemize}

\begin{table}[h!]
\centering
\begin{tabular}{|l|l|l|l|l|l|}
\hline
\textbf{Type} & \textbf{Input size} & \textbf{Digital Acc} & \textbf{Simulation Acc} & \textbf{Ratio} & \textbf{Ratio (\%)}\\
\hline
FC Simple Relu & 12x12 & 90.2 & 89.3 & 0.990 & 99.00\\
\hline
FC Simple Relu & 16x16 & 91.5 & 91.7 & 1.002 & 100.22\\
\hline
FC Double Relu & 12x12 & 91.9 & 90.3 & 0.982 & 98.25\\
\hline
FC Double Relu & 16x16 & 93.6 & 92.8 & 0.991 & 99.10\\
\hline
CV Simple Relu & 12x12 & 91.3 & 89.2 & 0.976 & 97.60\\
\hline
CV Simple Relu & 16x16 & 94.7 & 93.0 & 0.982 & 98.20\\
\hline
CV Double Relu & 12x12 & 91.5 & 88.0 & 0.961 & 96.10\\
\hline
CV Double Relu & 16x16 & 94.7 & 93.2 & 0.984 & 98.40\\
\hline
FC Double Sigmoid & 12x12 & 86.9 & 90.0 & 1.035 & 103.50\\
\hline
\end{tabular}
\caption{Accuracy results comparison between the digital network and the analog simulation}
\label{tab:results}
\end{table}

The results are very promising, always around a few percentage points of the digital network. In some cases, even surpassing the original; an unexpected behavior for which we might be able to offer a reasonable explanation.

The mistakes the simulated analog circuit makes are mostly due to the variability it was purposely built with. This variability will not affect the results for most inputs, i.e., the maximum value will correspond to the same output neuron even though its magnitude may vary. It will make mistakes where the digital network won't if the input makes the two output neurons with the highest activation values have very close values to each other.

On the other hand, when the digital network makes a mistake, the activation of an incorrect output neuron will have been the largest, but it is reasonable to assume that, if it is a high-quality neural network, the value of the output neuron with the correct answer was a close second. Now, given that our physical model has been built with an intrinsic variability by design, if we have a very close activation value on two outputs, the maximum of them might change with the variability. This would make the physical model sometimes able to make the right prediction where its digital counterpart will always fail.

\begin{figure}[h!]
	\centering
	\includegraphics[width =1\textwidth]{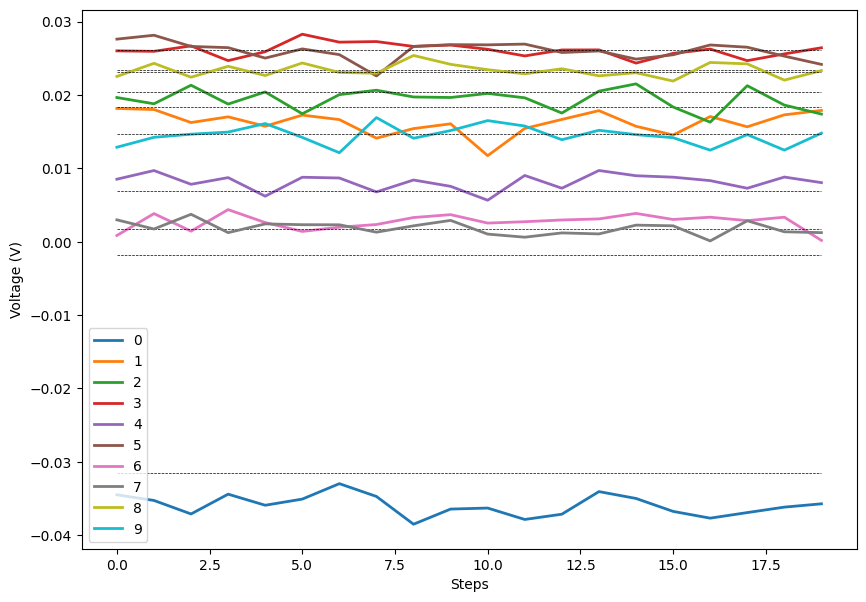}
	\caption{Values of the output neurons. Colored lines represent the outputs of the analog simulation, dashed lines the outputs of the digital neural network.}
 \label{fig:variability}
\end{figure}

To illustrate this phenomenon, we can select an example where the digital failed and the analog simulation succeeded,and see how the variability is responsible for this effect. to do this, we step $20$ times for the same input. The results are in figure \ref{fig:variability}

The right answer was $3$, but the digital predicted $5$. We can clearly see that the maximum between the neurons that represent $3$ and $5$ are constantly changing with the variability parameters.

Looking at the overall results, it is reasonable to assume that, in most cases, variability will lead to errors rather than the other way around.

Another interesting observation is that the performance of physical circuits improves relatively as the input size increases. In table \ref{tab:results} we can see that, for every architecture with two input size variants, the larger one has a slighter higher ratio. We would require more data to be certain, but this seems to suggest that the wider the network, the lesser the negative effects of the variability.

The depth of the network does not appear to affect the comparisons in either way, but again, more tests would be required. This would be an interesting finding, since it would mean we can scale these implementations with no considerable penalty to the performance.

\section{Conclusions and Future Work}
\label{cap:conclusiones}

We have successfully designed and simulated analog circuits based on memristors to perform inference on neural networks. For the simulations, we have used a state of the art physical model of the memristor that we were able to adjust to very reasonably match the behavior and empirical measurements of our real commercial device. We also managed to identify and overcome the restrictions our circuit design imposed over the parameters of the neural networks.

The results obtained are very satisfactory, as we have demonstrated that these analog circuits can obtain similar levels of performance.

It appears that there is limited literature available on neural networks built with Knowm devices that take into account variability. It's also noted that while some researchers \cite{florini2022hybrid} use the same devices but with different models and learning rules, they tend to use Spike-Timing-Dependent Plasticity (STDP) networks rather than convolutional networks. This could be due to the inherent properties of Knowm devices that may be more compatible with the principles of STDP, which is a biological process. However, the application of Knowm devices in convolutional networks could be an interesting area for future research, given the potential of these devices in mimicking synaptic behavior. 

The part of this work pertaining to memristor modeling has been published in the 14th Spanish Conference on Electron Devices (CDE 2023) \cite{cde2023characterization}

\section*{Acknowledgments}
We want to express our most sincere gratitude to our TFG advisors, Guillermo and Francisco, for their invaluable help and commitment to this project. Throughout almost a year of work, their guidance and support have been essential to the development and success of our work.

\bibliographystyle{unsrt}  
\bibliography{main.bbl}

\end{document}